\documentclass[11pt,a4paper]{article}
\usepackage{jheppub}
\usepackage{amsmath,amsfonts,amssymb}
\usepackage[T1]{fontenc}
\usepackage[utf8]{inputenc}
\usepackage[british]{babel}
\usepackage{lmodern}
\usepackage{microtype}
\usepackage{tikz}
\usepackage{slashed}
\usepackage{placeins}
\usepackage[compat=1.1.0]{tikz-feynhand}
\usepackage{comment}
\usepackage{xfrac}
\usepackage{color}
\usepackage{xcolor,colortbl}
\usepackage{xspace}
\allowdisplaybreaks

\usepackage{url}

\newcommand{\eminus}{\vcenter{\hbox{\scalebox{0.6}[1]{$ - $}}}}	
\newcommand{\bef}{$ \beta $-function\xspace}
\newcommand{\befs}{$ \beta $-functions\xspace}
\newcommand{\cofq}[2]{\mathbf{q}^{(#1)}_{#2}}
\newcommand{\Tr}[1]{\mathop{\mathrm{Tr}}\!\big[#1\big]\!}
\newcommand{\U}{\mathrm{U}}
\newcommand{\SU}{\mathrm{SU}}

\newcommand{\ud}[2]{\phantom{}^{#1}\phantom{}_{#2}}
 
\newcommand{\dd}{\mathop{}\!\mathrm{d}}

\newcommand{\eq}[1]{\eqref{eq:#1}}

\newcommand{\tab}[1]{Tab.~\ref{tab:#1}}
\newcommand{\Sec}[1]{Sec.~\ref{sec:#1}}
\newcommand{\App}[1]{App.~\ref{sec:#1}}
\newcommand{\tr}{\mathrm{tr}\!}

\newcommand{\mf}{\mathrm{m}}

\definecolor{blue}{rgb}{0,0.396,0.741}
\definecolor{MathBlue}{rgb}{0.368417, 0.506779, 0.709798}
\definecolor{MathYellow}{rgb}{0.880722, 0.611041, 0.142051}
\definecolor{MathGreen}{rgb}{0.560181, 0.691569, 0.194885}
\definecolor{MathRed}{rgb}{0.922526, 0.385626, 0.209179}
\definecolor{MathViolet}{rgb}{0.528488, 0.470624, 0.701351}

\begin{document}

\title{\boldmath General Quartic \texorpdfstring{$ \beta $-Function}{Beta Functions} at Three Loops}%
\abstract{
We determine the three-loop $\overline{\text{MS}}$ quartic $ \beta $-function for the most general renormalisable four-dimensional theories. A general parametrization of the $ \beta $-function is compared to known $ \beta $-functions for specific theories to fix all coefficients. Three-loop $ \beta $-functions for the cubic coupling and scalar mass terms also follow from the result, which is made available in software packages.    
}

\author[a]{Tom Steudtner,} 
\affiliation[a]{
	Fakultät für Physik, TU Dortmund, Otto-Hahn-Str. 4, D-44221 Dortmund, Germany
}
\emailAdd{tom2.steudtner@tu-dortmund.de}
\author[b]{Anders Eller Thomsen}
\emailAdd{anders.thomsen@unibe.ch}
\affiliation[b]{Albert Einstein Center for Fundamental Physics, Institute for Theoretical Physics, University of Bern, CH-3012 Bern, Switzerland}

\maketitle

\section{Introduction}

The renormalisation group (RG) is a deeply rooted feature of Quantum Field Theory (QFT). 
It is a consequence of physics being intrinsically dependent on the interaction scales and allows for evolving the theory according to the scale. Accordingly, many high-precision predictions from high-energy physics to critical phenomena rely on the availability of RG equations in advanced loop orders. 
This has inspired a long tradition of computing template RG equations: precomputed results so generic that they can be specialised to any renormalisable QFT,  
eliminating the need for lengthy and highly sophisticated calculations~\cite{Machacek:1983tz,Machacek:1983fi,Machacek:1984zw,Jack:1984vj,Pickering:2001aq,Luo:2002ti,Mihaila:2012pz,Sperling:2013eva,Sperling:2013xqa,Mihaila:2014caa,Schienbein:2018fsw}.
In practice, however, it has even proven challenging to evaluate the template RGs for particular model, which has prompted the development of an array of software packages~\cite{Staub:2013tta,Litim:2020jvl,Sartore:2020gou,Thomsen:2021ncy,Steudtner:FoRGEr} to facilitate the general results.

Generic results for the coupling \befs of renormalisable QFTs in the $\overline{\text{MS}}$ scheme~\cite{tHooft:1973mfk,Bardeen:1978yd} have recently become available 
up to four loops in the gauge sector and three loops for Yukawa interactions~\cite{Poole:2019kcm,Poole:2019txl,Bednyakov:2021qxa,Davies:2021mnc}.
However, the general RG equations for scalar coupling are available only up to two-loop order, as it has been for the last 40 years.
The limitation of available loop orders is foremost a consequence of the $\gamma_5$-problem that comes with the dimensional regularisation scheme~\cite{Jegerlehner:2000dz}. Higher loops require a more careful handling of this issue, which hampers progress due to the increased complexity of the calculation. 
Nevertheless, the drive for more precision has produced many partial results avoiding the problem. For instance, the issue is absent in purely scalar field theories, where six-loop \befs are available~\cite{Bednyakov:2021ojn}. In generic models with fermions and scalars but no gauge fields, three-loop RG equations were extracted \cite{Steudtner:2021fzs,Jack:2023zjt} from literature results and supersymmetry relations without performing a loop calculation. And yet, complete three-loop RG equations for all couplings in any renormalisable QFT has remained unavailable despite arguments brought forth in~\cite{Chetyrkin:2012rz,Poole:2019kcm,Litim:2023tym}, which show that $\gamma_5$ issues are absent in three-loop scalar \befs. 

The purpose of this work is to close that gap and compute three-loop quartic $\beta$-functions in a general formalism that covers all renormalisable QFTs. We follow the approach of fitting the template \befs to calculations in specific models, a strategy which has successfully been followed in~\cite{Steudtner:2020tzo,Steudtner:2021fzs,Bednyakov:2021qxa,Davies:2021mnc,Jack:2023zjt}. In addition to providing the analytic formulas, the results have also been implemented in the software packages \texttt{FoRGEr}~\cite{Steudtner:FoRGEr} and \texttt{RGBeta}~\cite{Thomsen:2021ncy}.
The general formalism is detailed in~\Sec{Formalism} while a short overview about the computational strategy is found in~\Sec{Computation}. The results are collected in~\Sec{Results} and a final outlook is given in~\Sec{Outlook}.

\section{Formalism}\label{sec:Formalism}
\paragraph{Notation.} In this work, we closely follow the notation developed in \cite{Steudtner:2021fzs}, which is mostly compatible with the one used in \cite{Poole:2019kcm}. We are interested in a generic QFT described by 
\begin{equation}\label{eq:master-template}
  \begin{aligned}
    \mathcal{L} =
     & \  \tfrac{1}{2} (D^{\mu})^{ca} \phi_a (D_\mu)^{cb} \phi_b + \tfrac{i}{2} \psi^i \tilde{\sigma}^\mu (D_\mu)_i^{\phantom{j}j} \psi_j  - \tfrac14 g^{\eminus 2}_{AB} \, F_{\mu\nu}^A F^{B\mu\nu} + \mathcal{L}_\text{gf} + \mathcal{L}_\text{gh}\\
     &  - \tfrac{1}{2} y^{ajk} \,\phi_a(\psi_j \varepsilon \psi_k)   - \tfrac{1}{24} \lambda^{abcd} \,\phi_a \phi_b \phi_c \phi_d\,\\
     &  - \tfrac{1}{2} \mf^{jk} \,(\psi_j \varepsilon \psi_k)  
-\tfrac{1}{2} (m^2)^{ab} \, \phi_a \phi_b - \tfrac{1}{6} h^{abc} \,\phi_a \phi_b \phi_c\,,
  \end{aligned}
  \end{equation}
  with the covariant derivatives
  \begin{equation}
    (D_\mu)_j^{\phantom{j}k} = \delta_j^{\phantom{j}k} \partial_\mu + i \,A^A_\mu (t^A)_j^{\phantom{j}k} \,, \qquad (D_\mu)_{ab} = \delta_{ab} \partial_\mu + i A^A_\mu T^A_{ab} \,,
  \end{equation}
  the field strength tensor
  \begin{equation}
    F^A_{\mu\nu} = \partial_\mu A^A_\nu - \partial_\nu A^A_\mu + f^{ABC} A^{B}_{\mu} A^{C}_{\nu} 
  \end{equation}
  as well as ghost and gauge-fixing terms $\mathcal{L}_\text{gh}$ and $\mathcal{L}_\text{gf}$, respectively.\footnote{The coupling \befs of the theory are known to be independent of the gauge-fixing terms, and our results are generally applicable.}  The scalar fields $\phi^a$ carry indices $a,b,c,\dots$ enumerating all real components. Hence, their raised or lowered positions do not carry meaning.
  The indices $i,j,k,\dots$ enumerate the distinct two-component Weyl fermions and their conjugates, both contained in $\psi^i$. Thus, $\psi^i$ as a tensor is pseudo-real, raising and lowering of its index amounts to complex conjugation $\psi^i = \psi_i^*$, which is, in fact, only a pairwise permutation of $\psi^i$.
  Both scalar and fermionic indices refer to different species, flavour and gauge indices altogether.
  Note that this construction implies a symmetry scalar quartic interactions $\lambda^{abcd} = \lambda^{(abcd)}$, Yukawa couplings $y^{aij} = y^{a(ij)}$, scalar cubics $h^{abc} = h^{(abc)}$, as well as fermionic $\mf{}^{ij}=\mf{}^{(ij)}$ and scalar masses $m^2_{ab} = m^2_{(ab)}$.
  Spinor indices of $\psi^i$ on the other hand are kept implicit here and are raised and lowered by the two-dimensional Levi-Civita $\varepsilon$. Moreover, $\tilde{\sigma}^\mu$ corresponds to $\sigma^\mu$ if it acts on a left-chiral Weyl fermion or $\bar{\sigma}^\mu$ otherwise. 
  Gauge fields carry the indices $A,B,C,\dots$ which both mark different gauge groups and also function as adjoint gauge indices, that is, they are the collection of adjoint indices from all product groups. 
  Perturbative calculations in this theory yields contractions of the tensor couplings in the second and third line of \eq{master-template}, as well as the fermionic and scalar generators $t^A_{ij}$ and $T^A_{ab}$ and the antisymmetric tensor $f^{ABC}$ from the gauge sector. The gauge indices $A,B,C,\dots$ of the latter two always appear raised and need to be contracted via the squared gauge couplings $g^2_{AB}$. This contraction is made implicit by lowering indices. The tensor $g^2_{AB}$ is block diagonal for non-Abelian gauge groups but can also accommodate gauge--kinetic mixing among Abelian subgroups (see \cite{Poole:2019kcm} for details).
  
  For convenience, the fermionic indices $i,j,k,\dots$ will also be made implicit as matrix multiplication when Yukawa or generator contractions are grouped together. For instance 
  \begin{equation}
    \tr{}\big[y^{acbc}t^{AB}\big] =  y^{a}_{ij} \,y^{c\,jk} \,y^{b}_{kl}\, y^{c\,lm} t^A_{mn} t^{B\,ni} 
  \end{equation}
  is understood. Moreover, shorthand notations for the Casimir 
  \begin{equation}
    (C_2^\mathrm{S})^a_{\phantom{a}b} = (T^A)^{a}_{\phantom{a} c} (T_{A})^{c}_{\phantom{c} b}\,\qquad (C_2^\mathrm{F})^i_{\phantom{i}j} = (t^A)^{i}_{\phantom{i} k} (t_{A})^{k}_{\phantom{k} j} \,, \qquad (C_2^\mathrm{G})^{AB} = f^{ACD}f^{B}_{\phantom{B}CD}
  \end{equation}
  and Dynkin tensors 
  \begin{equation}\label{eq:Dynkin}
    (S_2^\mathrm{S})^{AB} = \tr{}\,\big[T^A T^B\big] \,, \qquad (S_2^\mathrm{F})^{AB} = \tfrac12 \tr{}\,\big[t^A t^B\big] 
  \end{equation}
  are introducted. The factor $\tfrac12$ in $S_2^\mathrm{F}$ is purely conventional to connect to the definition of the fermionic Dynkin index in older works such as~\cite{Machacek:1983tz,Machacek:1983fi,Machacek:1984zw,Luo:2002ti,Schienbein:2018fsw}, where the trace would only sum over all Weyl fermions, whereas here the sum includes their complex conjugates. For convenience, we will also use the additional notation
  \begin{equation}
    Y_2^{ab} = \tr{}\,\big(y^a y^b\big)\,.
  \end{equation}
  
  In general, the $R_\xi$ gauge fixing requires the introduction of another tensor coupling $\xi^{AB}$ of gauge-fixing parameters. However, we are only interested in gauge-independent quantities here and omit this detail.

  \paragraph{Generic Tensor Structures.} The strategy of this work is the same as discussed in e.g.~\cite{Poole:2019kcm,Steudtner:2020tzo}: using the theory \eq{master-template}, ansätze for all $\beta$- and $\gamma$-functions can be formulated in terms of contractions of all interaction tensors discussed above, with model-independent coefficients.
  In particular, field anomalous dimensions $\gamma_{\alpha\alpha'}$ -- with $\alpha^{(\prime)}$ being any corresponding field indices -- are all rank 2 tensors, which is a sum containing all 1-particle irreducible contractions of tensor couplings with 2 open indices
  \begin{equation}\label{eq:TS-expansion}
    \gamma^{\alpha_1\alpha_2} = \sum_{n} c_n \, T_n^{\alpha_1\,\alpha_2\,\alpha_3\,\alpha_4\,\cdots\,\alpha_{i_n}} \delta_{\alpha_3 \,\alpha_4} \cdots \delta_{\alpha_{i_n-1}\,\alpha_{i_n}} \,.
  \end{equation}
  Here the tensors $T_n$ are direct products of interactions $y^a$, $\lambda^{abcd}$, $t^A$, $T^A$, $f^{ABC}$, $\mf{}$, $(m^2)^{ab}$, $h^{abc}$ from \eq{master-template}. Thus, the $T_n$ tensors inherit all the model-dependent information from these couplings: their dimensions reflect the field content of the theory, while their components specify all manifest symmetries and interactions. On the other hand, the coefficients $c_n$ are model independent; they are determined from resolving the Lorentz and spinor algebra and extracting the counterterms of the relevant Feynman diagrams. 
  Each coupling $X$ at an $N$-point vertex has a \bef of the shape 
  \begin{equation}\label{eq:beta-ansatz}
    \beta_X^{ \alpha_1\, \alpha_2 \, \dots \,\alpha_N} = \hat{\gamma}^{\alpha_1}_{\phantom{\alpha_1}\alpha'} X^{\alpha' \, \alpha_2 \,\dots \,\alpha_N} + \dots + \hat{\gamma}^{\alpha_N}_{\phantom{\alpha_N}\alpha'} X^{\alpha_1 \, \alpha_2 \,\dots \,\alpha'} + \tilde{\beta}_X^{ \alpha_1\, \alpha_2 \, \dots \,\alpha_N}
  \end{equation}
  which consists of both leg $\hat{\gamma}$ and vertex corrections $\tilde{\beta}_X$. The latter may be expanded in 1-particle irreducible tensor of rank $N$ as in \eq{TS-expansion}. Note that the leg corrections $\hat{\gamma}$ may have an identical basis of tensor structures to field anomalous dimensions $\gamma$, but not necessarily the same coefficients.
  The reason for this is the gauge symmetry relations~\cite{Jack:2014pua,Poole:2019kcm} 
  \begin{equation}\label{eq:gauge-relations}
    \begin{aligned}
      0 &= t^A y^a - y^a t^A +  y^b (T^A)^{ba}\,, \\
      0 &= (T^A)^{ae} \lambda^{ebcd} + (T^A)^{be} \lambda^{eacd} + (T^A)^{ce} \lambda^{eabd} + (T^A)^{de} \lambda^{eabc} \\
      0 &= [t^A, t^B] - i f^{ABC} t^C \,,\\
      0 &= [T^A, T^B]^{ab} - i f^{ABC} (T^C)^{ab}\,, \\
      0 &= f^{ABE} f^{CDE} - f^{ACE} f^{BDE} + f^{ADE} f^{BCE}\,,\,
    \end{aligned}
  \end{equation}
  and similar ones for the super-renormalisable couplings. These relations represent a freedom of transforming coefficients $c_n$ of tensor structures without changing the result of the RGEs. This allows for greatly reducing the basis of tensor structures.
  It also indicates that the basis choice is not unique.
  Curiously, gauge relations \eq{gauge-relations} may bring  leg corrections into the shape of vertex corrections and vice versa. This accounts for the distinction of  $\hat{\gamma}$ and  $\gamma$ in \eq{TS-expansion} and \eq{beta-ansatz}.
  In fact, while both vertex and leg corrections may depend on the gauge-fixing, the \bef is independent of the gauge. This cancellation is achieved at the level of tensor structures by the means of gauge relations.

\section{Computation}\label{sec:Computation}

The procedure to obtain a three-loop scalar quartic \bef for the generic renormalisable QFT is as follows:
Using the formalism described in \Sec{Formalism}, we utilise the \texttt{GRAFER} code~\cite{Poole:2019kcm} to generate a basis of tensor structures for the \bef in question.
These terms are then implemented in the symbolic RG codes \texttt{RGBeta}~\cite{Thomsen:2021ncy} and \texttt{FoRGEr}~\cite{Steudtner:FoRGEr}, which allows for a cross-check of their correctness.
The tensor structures are implemented with symbolic variables
serving as placeholders for the model-independent numeric coefficients multiplying each tensor structure in the quartic \bef.
These coefficients are determined by choosing example QFTs, computing the \befs both with the RG codes~\cite{Thomsen:2021ncy,Steudtner:FoRGEr} and explicit loop calculations. Comparing terms featuring the same couplings and quantum numbers yields constraints on and, eventually, fixes all unknown coefficients.

In our case, the required three-loop calculations were conducted with a custom version of the framework \texttt{MaRTIn}~\cite{Brod:2024zaz}, employing infrared rearrangement via a mass parameter~\cite{Misiak:1994zw,Chetyrkin:1997fm} to isolate UV from IR poles.
The $\gamma_5$ problem~\cite{Jegerlehner:2000dz} can be circumvented for the three-loop quartic \bef as shown in~\cite{Litim:2023tym,Poole:2019kcm,Chetyrkin:2012rz}. It turns out that there is no reading-point ambiguity and thus the semi-na\"ive $\gamma_5$ scheme~\cite{Chetyrkin:2012rz,Bednyakov:2012en}  provides a consistent result and is adopted here.
The computational setup has already been utilised in \cite{Litim:2023tym} where cross-checks with the literature were made. We have also reproduced the three-loop SM Higgs quartic \bef with general Yukawa structure~\cite{Chetyrkin:2012rz,Bednyakov:2013eba,Chetyrkin:2013wya,Bednyakov:2013cpa,Bednyakov:2014pia} as a non-trivial test of \texttt{MaRTIn}.
Furthermore, consistency of all constraints on the numerical \bef coefficients across all quartic couplings and models are a highly non-trivial consistency check.

In the following, we introduce several example models which we have used to extract the fully general \bef. We do not know of any good way to guarantee from the outset that the different models do not lead to degenerate conditions on the coefficients. Instead, we have tried to come up with a selection that exhibit a variety of different quartic couplings in rich models that might break degeneracies between the different tensor structures. As in previous implementations of the strategy a handful of models have proven sufficient. 
In fact the two gauged models (secs.~\ref{sec:lq_model} and~\ref{sec:adjoint_model}) are sufficient to fix all unknown coefficients and are consistent with each other while exhibiting significant overlap on the constraints for most coefficients. Moreover, our findings are in accord with independent literature results such as the SM quartic~\cite{Chetyrkin:2012rz,Bednyakov:2013eba,Chetyrkin:2013wya,Bednyakov:2013cpa,Bednyakov:2014pia}, Abelian Higgs~\cite{Ihrig:2019kfv} and Abelian Gross-Neveu-Yukawa models~\cite{Zerf:2020mib}, the Litim-Sannino model~\cite{Litim:2023tym} and the gaugeless inputs collected in~\cite{Steudtner:2021fzs,Jack:2023zjt}. 
Before getting to the models, we need to discuss an important ambiguity in the quartic \bef.

\subsection{\texorpdfstring{$ \boldsymbol{\beta} $}{Beta}-Function Ambiguity}\label{sec:beta-ambiguity}
It has been established that RG functions suffer from an ambiguity associated with unphysical rotations in the flavour group of the matter fields~\cite{Herren:2021yur}. This was first observed for the non-Abelian flavour (generation) symmetries of the SM~\cite{Bednyakov:2012en,Herren:2017uxn}, but generally occur, starting at three-loop order, whenever the couplings of the theory are non-trivial in flavour space. To see the origin of the ambiguity consider the the wave-function renormalisation of a scalar kinetic term\footnote{For real scalars as in Eq.~\eqref{eq:master-template}, the following discussion adapts trivially by using transpose rather than Hermitian conjugate and anti-symmetric rather than anti-Hermitian. Similar considerations hold for the fermion wave-function renormalisation constants too.} 
	\begin{equation}
	\mathcal{L} \supset (Z_\phi^\dagger Z_\phi^{\phantom{\dagger}})\ud{a}{b} \,D_\mu \phi_a^\ast \, D^\mu \phi^b.
	\end{equation} 
It is somewhat non-standard to factorize the renormalisation constants by $ Z_\phi^\dagger Z_\phi^{\phantom{\dagger}} $ rather than single, Hermitian matrix $ Z \equiv Z_\phi^\dagger Z_\phi^{\phantom{\dagger}} $, but this factorization is in some sense more honest; it more clearly elucidates the ambiguity in the RG functions. In the $ \overline{\text{MS}} $ scheme, finiteness of the 2-point function fixes the combination $ Z_\phi^\dagger Z_\phi^{\phantom{\dagger}} $ uniquely. However, there can still be freedom in choosing $ Z_\phi^{\phantom{\dagger}} $ that lead to the same $ Z_\phi^\dagger Z_\phi^{\phantom{\dagger}} $.
As $ Z_\phi^{\phantom{\dagger}} $ determines the field anomalous dimension
	\begin{equation}
	\gamma_\phi = Z_\phi^{\eminus 1} \dfrac{\dd}{\dd \,\ln \mu} Z_\phi^{\phantom{\dagger}},
	\end{equation}
the ambiguity in $ Z_\phi $ transfers into the anomalous dimensions and the \befs too, as changes in $ Z_\phi $ has to be compensated by appropriate changes in the coupling counterterms.

Our calculation of the RG functions of the model in Sec.~\ref{sec:lq_model} is the first observation of the ambiguity for matter fields with an Abelian flavour symmetry (though, see also~\cite{Fortin:2012cq}). The complex scalars of this theory each posses a $ \U(1) $ flavour symmetry, which allows the scalar anomalous dimensions to gain a phase. Beginning at three-loop order there are three tensor structures 
\setlength{\feynhanddotsize}{1mm} 
\begin{equation}\label{eq:asymmetric-legs}
    \begin{tikzpicture}[baseline=-0.25em]
      \begin{feynhand}
        \vertex [dot] (l0) at (0em,0em) {};
        \vertex [dot] (l1) at (-2em,0em);
        \vertex (c0) at (2em,0em);
        \vertex [dot] (c1) at (2em,2em) {};
        \vertex [dot] (c2) at (2em,-2em) {};
        \vertex [dot] (r0) at (4em,0em)  {};
        \vertex [dot] (r1) at (6em,0em);
        \vertex [dot] (u1) at (0.11em,0.6em) {};
        \vertex [dot] (u2) at (1.4em,1.9em) {};
        \propag [scalar] (l0) to (l1);
        \propag [scalar] (r0) to (r1);
        \propag [scalar] (c1) to (c2);
        \propag [photon] (u1) to [half right, looseness=1.5] (u2);
        \draw (c0) circle (2em);
      \end{feynhand}
    \end{tikzpicture} \qquad
    \begin{tikzpicture}[baseline=-0.25em]
      \begin{feynhand}
        \vertex [dot] (l0) at (0em,0em) {};
        \vertex [dot] (l1) at (-2em,0em);
        \vertex [dot] (c0) at (2em,0em) {};
        \vertex [dot] (c1) at (2em,2em) {};
        \vertex [dot] (c2) at (2em,-2em) {};
        \vertex [dot] (r0) at (4em,0em)  {};
        \vertex [dot] (r1) at (6em,0em);
        \propag [scalar] (l0) to (l1);
        \propag [scalar] (r0) to (r1);
        \propag [plain] (c1) to [half right, looseness=1.6 ] (c2);
        \propag [plain] (c1) to (c2);
        \propag [scalar] (c1) to [half left, looseness=1.6 ] (c2);
        \propag [scalar] (c0) to (r0);
      \end{feynhand}
    \end{tikzpicture} \qquad
    \begin{tikzpicture}[baseline=-0.25em]
      \begin{feynhand}
        \vertex [dot] (l0) at (0em,0em) {};
        \vertex [dot] (l1) at (-2em,0em);
        \vertex (c0) at (2em,0em);
        \vertex [dot] (c1) at (2em,2em) {};
        \vertex [dot] (c2) at (2em,-2em) {};
        \vertex [dot] (r0) at (4em,0em)  {};
        \vertex [dot] (r1) at (6em,0em);
        \vertex [dot] (u1) at (0.11em,0.6em) {};
        \vertex [dot] (u2) at (1.4em,1.9em) {};
        \propag [scalar] (l0) to (l1);
        \propag [scalar] (r0) to (r1);
        \propag [scalar] (c1) to (c2);
        \propag [scalar] (u1) to [half right, looseness=1.5] (u2);
        \draw (c0) circle (2em);
      \end{feynhand}
    \end{tikzpicture}
\end{equation}
contributing to $ \gamma_\phi $ or equivalenty $ Z_\phi $ that allow for anti-Hermitian, or in the Abelian case simply imaginary, combinations~\cite{Poole:2019kcm}.\footnote{The coefficients of the template expansion~\eqref{eq:TS-expansion} has to be real to accommodate the case of real scalars; they cannot be the source of any phase.} 
The coefficients multiplying these anti-Hermitian combinations are generally entirely unconstrained by the renormalisation condition. 
We fix the ambiguity by enforcing that $ Z_\phi $ is Hermitian in all cases. This coincides with the choice used in~\cite{Davies:2021mnc} for the three-loop Yukawa \bef, crucially ensuring compatibility of the two results. Following this convention, the flavour-improved \bef for the quartic coupling
\begin{equation}
   B^{abcd}_\lambda = \beta^{abcd}_\lambda - \upsilon^{ae}\, \lambda^{ebcd} - \upsilon^{bd}\, \lambda^{eacd} - \upsilon^{ce}\, \lambda^{eabd} - \upsilon^{de}\, \lambda^{eabc}
\end{equation}
can be obtained by using the RG function $ \upsilon $~\cite{Davies:2021mnc} in addition to the results we present here. 

A final interesting feature of the flavour ambiguity in the case of an Abelian flavour group is that it does not lead to divergent RG functions at the three-loop order when $ Z_\phi $ is chosen Hermitian. This is in contrast to non-Abelian flavour groups, where one has to carefully engineer $ Z_{\phi,\psi} $ to avoid (otherwise harmless) $ \epsilon $-poles in the RG functions. On general grounds, for instance, the choice of fixing $ Z_{\phi,\psi} $ to be Hermitian leads exactly to such poles~\cite{Herren:2021yur}. In contrast to the non-Abelian symmetries, the absence of any poles in the Abelian case serves to partially obscure the RG ambiguity, and it might easily be overlooked.

\subsection{Gaugeless Model}
In the gaugeless limit, a basis for all $\beta$- and $\gamma$-functions was developed in~\cite{Steudtner:2021fzs}. Scalar and fermion anomalous dimensions as well as Yukawa \befs could all be completely determined by an earlier loop calculation in the THDM~\cite{Herren:2017uxn}. For three-loop quartic vertex corrections on the other hand, the data was insufficient. 
However, the basis was completed in~\cite{Jack:2023zjt} using the pure scalar \befs~\cite{Jack:1990eb}, the Higgs self-coupling in the SM with $g_1 = g_2 = g_3 = 0$~\cite{Chetyrkin:2012rz,Bednyakov:2013eba,Chetyrkin:2013wya,Bednyakov:2013cpa,Bednyakov:2014pia}, results from Gross-Neveu-Yukawa theories~\cite{Zerf:2017zqi,Mihaila:2017ble}, $\mathcal{N}=1$ supersymmetry relations and explicit results in the $\overline{\text{DR}}$ scheme with vanishing gauge coupling~\cite{Jack:1996qq,Parkes:1985hh}, as well as $\mathcal{N}=\tfrac12$ supersymmetry relations~\cite{Fei:2016sgs,Jack:2023zjt}.

While the 62 coefficients of the gaugeless theory were over-constrained by the references above, fixing them by a single computation has not been achieved previously.
In this work, we have computed the complete three-loop vertex corrections to the quartic \befs of the theory
\begin{equation}
  \mathcal{L} = \overline{\Psi}_i i \slashed{\partial} \Psi_i  + \tfrac12 \partial_\mu\Phi_a \partial^\mu\Phi_a - Y^{aij} \Phi_a\overline{\Psi}_i \Psi_j - \tfrac1{4!} \lambda^{abcd} \Phi_a \Phi_b \Phi_c \Phi_d\,,
\end{equation}
featuring $N_\Psi$ Dirac fermions $\Psi_i$ and $N_\Phi$ real scalars $\Phi_a$. 
The Yukawa and quartic couplings act as spurions of the $\U(N_\Psi) \times \mathrm{O} (N_\Phi)$ global flavour symmetry, which at three-loop order gives rise to an ambiguity of the RGs~\cite{Herren:2021yur}. Restricting ourselves to the vertex correction of the quartic coupling allows us to avoid these redundancies. 
The model is sufficient to fix all coeffcients in the gaugeless limit.

\subsection{SM \& Seesaw I \texorpdfstring{$\oplus$}{+} III \& 2 Leptoquarks }\label{sec:lq_model}

\begin{table}[t]
  \centering
  \begin{tabular}{|l|llll|}
    \hline
    Field & Gen. & $\U(1)_Y $ & $\SU(2)_L$ & $\SU(3)_c$ \\
    \hline 
    $Q$ & 3 & $+\tfrac16$ & $\mathbf{2}$ & $\mathbf{3}$ \\
    $U$ & 3 & $+\tfrac23$ & $\mathbf{1}$ & $\mathbf{3}$ \\
    $D$ & 3 & $-\tfrac13$ & $\mathbf{1}$ & $\mathbf{3}$ \\
    $L$ & 3 & $-\tfrac12$ & $\mathbf{2}$ & $\mathbf{1}$ \\
    $E$ & 3 & $-1$ & $\mathbf{1}$ & $\mathbf{1}$ \\
    $N$ & $N_1$ & $\phantom{+}0$ & $\mathbf{1}$ & $\mathbf{1}$ \\
    $S$ & $N_3$ & $\phantom{+}0$ & $\mathbf{3}$ & $\mathbf{1}$ \\
    \hline
    $H$ & 1 & $+\tfrac12$ & $\mathbf{2}$ & $\mathbf{1}$ \\
    $\phi$ & 1 & $+\tfrac16$ & $\mathbf{2}$ & $\mathbf{3}$ \\
    $\chi$ & 1 & $+\tfrac76$ & $\mathbf{2}$ & $\mathbf{3}$ \\
    \hline
  \end{tabular}
  \caption{Overview of left- ($Q$, $L$) and right-handed ($U$, $D$, $E$, $N$, $S$) Weyl fermions and complex scalars ($H$, $\phi$, $\chi$) as well as their representations under the $\U(1)_Y  \times \SU(2)_L \times \SU(3)_c$ gauge symmetry.}
  \label{tab:SM-I-III-LQ2}
\end{table}

As a second model to fix the coefficients of the generic quartic \bef, we employ a leptoquark model,  
the field content of which is summarised in \tab{SM-I-III-LQ2}. The theory features the Standard Model (SM) gauge group $\U(1)_Y \times \SU(2)_L \times \SU(3)_c$ and field content. In addition, there are two new types of right-handed fermions $N$ and $S$, which are $\SU(2)_L$ singlets and triplets with their own Yukawa interactions with the SM Higgs $H$, as featured in the type I and III seesaw mechanisms. Moreover, two scalar leptoquarks $\phi$ and $\chi$, $\SU(2)_L$ doublets with hypercharges $\tfrac16$ and $\tfrac76$, respectively, are introduced. This charge assignment prevents scalar mixing and gives rise to a rich Yukawa sector:
\begin{equation}
  \begin{aligned}
    - \mathcal{L}_\text{yuk} &= Y_d^{ij} Q^\dagger_i H D_j + Y_u^{ij} Q^\dagger_i \tilde{H} U_j + Y_e^{ij} L^\dagger_i H E_j + Y_n^{ij} L^\dagger_i \tilde{H} N_j + Y_s^{ij} L^\dagger_i \tilde{H} S_j  \\
    &{} \quad +  X_n^{ij} Q^\dagger_i \phi N_j +  X_s^{ij} Q^\dagger_i \phi S_j + X_d^{ij} L^\dagger_i \tilde{\phi} D_j +  X_e^{ij} Q^\dagger_i \chi E_j + X_u^{ij} L^\dagger_i \tilde{\chi} U_j + \text{H.c.}\,.
  \end{aligned}
\end{equation}
Here $\tilde{H} = \varepsilon H^\ast$ and $\varepsilon$ is the Levi-Civita tensor contracting $\mathbf{2} \otimes \mathbf{2}$ of the $SU(2)_L$ symmetry group.
The quartic couplings similarly feature a variety of different interactions:
\begin{equation}
  \begin{aligned}
    - \mathcal{L}_\text{quartic} &= \lambda (H^\dagger H)^2   + v \left(\tr{}\,\,\phi^\dagger \phi\right)^2  + u \,\tr{}\,(\phi^\dagger \phi \phi^\dagger \phi) + t \left(\tr{}\,\,\chi^\dagger \chi\right)^2   + s \,\tr{}\,(\chi^\dagger \chi \chi^\dagger \chi)
    \\
    &{}\ + \delta_1 \,(H^\dagger H) \tr{}\,(\phi^\dagger \phi) + \delta_2 \,( H \phi^\dagger \phi H^\dagger) + \delta_3 \,(H^\dagger H) \tr{}\,(\chi^\dagger \chi) \\
    &{}\ + \delta_4 \,( H \chi^\dagger \chi H^\dagger) + \delta_5 \,\tr{}\,(\phi^\dagger \phi) \tr{}\,(\chi^\dagger \chi) + \delta_6 \,\tr{}\,(\phi^\dagger \chi) \tr{}\,(\chi^\dagger \phi)  \\
    &{}\  + \delta_7 \,\tr{}\,(\phi^\dagger \phi \chi^\dagger \chi )  + \delta_8 \,\tr{}\,(\chi^\dagger \phi \phi^\dagger \chi ) + q\,H \chi^\dagger \phi \varepsilon H  + q^* H^\dagger \chi \phi^\dagger \varepsilon H^\dagger\, ,
  \end{aligned} 
\end{equation}
where we have implicitly written the leptoquarks $\phi$ and $\chi$ as matrices $\phi_{c\ell}$ with $c$ the colour and $\ell$ the weak isospin index. An auxiliary file with \befs for all quartic interactions is included with the arXiv version of this work.

The multiple generations of the fermion fields indicates the presence of a non-Abelian flavour symmetry (of the fermion kinetic terms), under which the Yukawa couplings act as spurions. As in the SM, this gives an ambiguity in the fermion anomalous dimension and and the Yukawa coupling \befs, starting at the three loop order~\cite{Herren:2021yur}. There are no duplicate representations of the scalar fields, and so their flavour group is Abelian: $ \U(1)_H \times \U(1)_{\phi} \times \U(1)_\chi $. The rich coupling structure gives rise to additional physical phases beyond the CKM phase of the SM. This allows for the presence of anti-Hermitian (imaginary) contributions to the scalar anomalous dimensions already at three-loop order.\footnote{The lowest-order invariant combination of the SM Yukawa matrices that is sensitive to CP--violation, and, thus, the physical phase, is the Jarlskog invariant, which cannot be generated perturbatively at three-loop order.} For instance, the Higgs anomalous dimension has an ambiguous contribution   
	\begin{equation}
	\gamma_H \subset \kappa\, \mathrm{tr} \big( Y_e^{\phantom{\dagger}}\! Y_e^\dagger \big[Y_n^{\phantom{\dagger}}\! Y_n^\dagger ,\,  Y_s^{\phantom{\dagger}}\! Y_s^\dagger \big] \big)  
	\end{equation}
(among others), where $ \kappa $ is not determined by the renormalisation of the theory. Our choice of using Hermitian scalar renormalisation constants fixes $ \kappa =0 $. 

This SM extension is very potent for fixing the coefficients of the quartic \bef: it provides 311 independent conditions on the 315 coefficients of the three-loop quartic \bef (228 conditions on the 232 coefficients remaining after fixing the gaugeless ones).

\subsection{Adjoint Model} \label{sec:adjoint_model}

\begin{table}[t]
  \centering
  \begin{tabular}{|l|ccc|}
    \hline
    Field & Gen. & $\SU(N_1) $ & $\SU(N_2)$  \\
    \hline 
    $Q$ & $N_Q$ & $N_1$ & $N_2$  \\
    $U$ & $N_U$ & $N_1$ & $N_2$  \\
    $D$ & $N_D$ & $N_1$ & $N_2$  \\
    $L$ & $N_L$ & $1$ & $N_2$  \\
    $E$ & $N_E$ & $1$ & $N_2$  \\
    $N$ & $N_N$ & $1$ & $N_2$  \\
    \hline
    $\phi_1$ & 1 & $1$ & $N_2^2-1$  \\
    $\phi_2$ & 1 & $1$ & $N_2^2-1$  \\
    $\phi_3$ & 1 & $N_1^2-1$ & $1$  \\
    \hline
  \end{tabular}
  \caption{Overview of Dirac fermions $Q$, $U$, $D$, $L$, $E$, $N$ and real scalar fields $\phi_{1,2,3}$ as well as their multiplicities under the $\SU(N_1) \times \SU(N_2)$ gauge group.}
  \label{tab:adjoint-model}
\end{table}

We have also used another toy model featuring an $\SU(N_1)  \times \SU(N_2)$ gauge group and vector-like fermions $Q$, $U$, $D$, charged in the fundamental under both subgroups, as well as $L$, $E$ and $N$ which are only fundamental under $\SU(N_2)$ but $\SU(N_1)$ singlets.
Moreover, three real scalar fields $\phi_{1,2,3}$ are included, each transforming as an adjoint under one of the product groups. The field content is summarised in \tab{adjoint-model}.
The Yukawa interactions are given by 
\begin{equation}
  \begin{aligned}
    - \mathcal{L}_\text{yuk} &=   Y_D^{ij} \bar{Q}_i \phi_1 U_j + Y_N^{ij} \bar{L}_i \phi_1 E_j  +  Y_U^{ij} \bar{Q}_i \phi_2 D_j + Y_E^{ij} \bar{L}_i \phi_2 N_j  +  Y_Q^{ij} \bar{U}_i \phi_3 D_j  + \text{H.c.} \,,
  \end{aligned}
\end{equation}
and the scalar potential reads
\begin{equation}
  \begin{aligned}
    V &=   \frac{\lambda_{11A}}2 \left(\tr{} \,\,\phi_1^2\right)^2 +  \frac{\lambda_{11B}}8 \tr{} \,\big(\phi_1^4\big) + \frac{\lambda_{22A}}2 \left(\tr{} \,\,\phi_2^2\right)^2 +  \frac{\lambda_{22B}}8 \tr{} \,\big(\phi_2^4\big)  \\
    &{}\, + \frac{\lambda_{33A}}2 \left(\tr{} \,\,\phi_3^2\right)^2 +  \frac{\lambda_{33B}}8 \tr{}\, \big(\phi_3^4\big)  + \lambda_{12A} \,\tr{} \, \big(\phi_1^2\big)\,\tr{}\,\big(\phi_2^2\big) + \lambda_{12B} \,\big(\tr{} \,\,\phi_1 \phi_2\big)^2 \\
    &{}\, + \frac{\lambda_{12C}}2 \,\tr{}\, \big(\phi_1^2 \phi_2^2\big) + \frac{\lambda_{12D}}4 \,\tr{}\, \big(\phi_1 \phi_2\big)^2 +  \lambda_{13} \,\tr{}\, \big(\phi_1^2\big) \tr{}\, \big(\phi_3^2\big) +  \lambda_{23} \,\tr{}\, \big(\phi_2^2\big) \tr{}\, \big(\phi_3^2\big)\,.
  \end{aligned}
\end{equation}
The model features three discrete symmetries
\begin{equation}\label{eq:discrete-symmetry}
  \begin{aligned}
    (\phi_1, \phi_2, \phi_3,  Q, U, D, L, E, N)  &\mapsto (-\phi_1, -\phi_2,  \phi_3, -i Q, i U, i D, -i L, i E , i N), \\
    (\phi_1, \phi_2, \phi_3,  Q, U, D, L, E, N)  &\mapsto (\phi_1, -\phi_2,  -\phi_3, i Q, i U, -i D, i L, i E , -i N), \\
    (\phi_1, \phi_2, \phi_3,  Q, U, D, L, E, N)  &\mapsto (-\phi_1, \phi_2,  -\phi_3, i Q, -i U, i D, i L, -i E , i N),
  \end{aligned}
\end{equation}
that prevent scalar mixing between the fields $\phi_1$ and $\phi_2$ and prevent additional Yukawa and quartic couplings from being induced in the RG evolution. Thereby the symmetry also prevents any anti-symmetric contributions to the $ \gamma $-function between them. Hence, there is no RG ambiguity in this model.

By itself, the model yields 166 conditions among the 232 non-gaugeless coefficients.

\section{Results}\label{sec:Results}

The general shape of a quartic \bef
\begin{equation}\label{eq:def-beta-lambda}
  \beta_\lambda^{abcd} = \hat{\gamma}^{ae}_\phi \lambda^{ebcd} + \hat{\gamma}^{be}_\phi \lambda^{eacd} + \hat{\gamma}^{ce}_\phi \lambda^{eabd} +  \hat{\gamma}^{de}_\phi \lambda^{eabc} +  \tilde{\beta}_\lambda^{abcd} 
\end{equation}
contains both leg corrections via the tensor structures contained in $\hat{\gamma}_\phi$ and vertex contributions $\tilde{\beta}_\lambda$. 
After applying gauge relations, a basis for the three-loop leg corrections reads 
\begin{equation}\label{eq:ad-general}
    \hat{\gamma}_\phi^{(3) \,ab}  =  \gamma_\phi^{(3) \,ab}  \Big|_{g=0}  + \gamma_{ g^2 y^4}^{(3) \,ab} + \gamma_{ g^4 y^2}^{(3) \,ab} + \gamma_{ g^2 \lambda^2}^{(3) \,ab} + \gamma_{ g^4 \lambda}^{(3) \,ab}  + \gamma_{ g^6 }^{(3) \,ab}\,. 
  \end{equation}
The gaugeless parts are simply given by the field anomalous dimensions listed in~\cite{Steudtner:2021fzs,Jack:2023zjt}. The remaining terms read

\begin{align}
&\begin{aligned}
  \gamma_{g^2 y^4}^{(3) \,ab} = &{} \ \gamma_1\,(Y_2 Y_2 C_2^\mathrm{S})^{ab} + \gamma_2\, \tr{}\left( y^{abc} C_2^\mathrm{F} y^c \right) + \gamma_3 \, \tr{}\left( y^{accb}  C_2^\mathrm{F} \right) \\
  &{}\  + \gamma_4 \, \tr{}\left( y^{abcc}  C_2^\mathrm{F} \right) + \gamma_5 \, (C_2^\mathrm{S})^{ac} \,\tr{}\left( y^{cbdd} \right) + \gamma_6\,\tr{}\left( y^{acdb} \right) \,(C_2^\mathrm{S})^{cd} \\
  &{}\  + \gamma_7\, \tr{}\left( y^a C_2^\mathrm{F} y^{cbc} \right) + \gamma_7\, \tr{}\left( y^{ac}  C_2^\mathrm{F} y^{bc} \right) + \gamma_8 \, (C_2^\mathrm{S})^{ac} \,\tr{}\left( y^{dcdb} \right) \\
  &{}\   + \gamma_{9}\,\tr{}\left( y^{acbd} \right) (C_2^\mathrm{S})^{cd}  +  \gamma_{10}\, \tr{}\left( y^{ac} t^A y^{cb} t_A \right) \,,
\end{aligned} \label{eq:ad-g2y4} \\[1em]
&\begin{aligned}
  \gamma_{g^2 \lambda^2}^{(3) \,ab} =  &{} \ \gamma_{11}\,\lambda^{acde} (C_2^\mathrm{S})^{cf} \lambda^{bdef} + \gamma_{12}\,(C_2^\mathrm{S})^{ac} \,\lambda^{cdef}\lambda^{bdef}\,,
\end{aligned}\label{eq:ad-g2l2}\\[1em]
&\begin{aligned}
  \gamma_{g^4 y^2}^{(3) \,ab} =  &{} \ \gamma_{13}\,\left(Y_2 T^{AB}\right)^{ab} (S_2^\mathrm{F})_{AB} +  \gamma_{14}\, \left(Y_2 T^{AB}\right)^{ab} (S_2^\mathrm{S})_{AB} \\
  &{}\ +  \gamma_{15}\,\left(Y_2 T^{AB}\right)^{ab} (C_2^\mathrm{G})_{AB} 
   +  \gamma_{16}\,\tr{}\left( y^a t^{AB} y^b \right) (S_2^\mathrm{F})_{AB} \\
   &{}\ + \gamma_{17}\,\tr{}\left( y^a t^{AB} y^b \right) (S_2^\mathrm{S})_{AB} 
   + \gamma_{18}\,\tr{}\left( y^a t^{AB} y^b \right) (C_2^\mathrm{G})_{AB} \\
   &{}\ + \gamma_{19}\, (C_2^\mathrm{S} Y_2 C_2^\mathrm{S})^{ab} + \gamma_{20}\,\tr{}\left( y^{ab} C_2^\mathrm{F} C_2^\mathrm{F}\right) + \gamma_{21}\,\tr{}\left( y^a C_2^\mathrm{F}  y^b C_2^\mathrm{F}\right) \\
   &{}\ + \gamma_{22}\, (C_2^\mathrm{S})^{ac}\,\tr{}\left( y^{bc} C_2^\mathrm{F}  \right) + \gamma_{23}\, (T^{AB})^{ab} \tr{}\left(T^{BA} Y_2\right) \\
   &{}\ + \gamma_{24}\,  (T_{AB})^{ab} \tr{}\left(t^{BA} y^{cc}\right)\,,
\end{aligned}\label{eq:ad-g4y2}\\[1em]
&\begin{aligned}
  \gamma_{g^4 \lambda}^{(3) \,ab} =  &{} \ \gamma_{25}\,\lambda^{acde} (T^{AB})^{cd} (T_{BA})^{eb} \,,
\end{aligned}\label{eq:ad-g4l}\\[1em]
&\begin{aligned}
  \gamma_{g^6}^{(3) \,ab} =  &{} \ \gamma_{26}\,\left( T^{AB} \right)^{ab} (S_2^\mathrm{F} S_2^\mathrm{F} )_{AB} + \gamma_{27}\,\left( T^{AB} \right)^{ab} (C_2^\mathrm{G} S_2^\mathrm{F} )_{AB} \\
  &{} +  \gamma_{28}\,\left( T^{AB} \right)^{ab} (S_2^\mathrm{F} S_2^\mathrm{S} )_{AB}  + \gamma_{29}\,\left( T^{AB} \right)^{ab} (C_2^\mathrm{G} C_2^\mathrm{G}  )_{AB} \\
  &{} + \gamma_{30}\,\left( T^{AB} \right)^{ab} (C_2^\mathrm{G} S_2^\mathrm{S}  )_{AB} +  \gamma_{31}\,\left( T^{AB} \right)^{ab} (S_2^\mathrm{S} S_2^\mathrm{S} )_{AB} \\ 
  &{} + \gamma_{32}\,\left( C_2^\mathrm{S} T^{AB} \right)^{ab} (S_2^\mathrm{F} )_{AB} + \gamma_{33}\,\left( C_2^\mathrm{S} T^{AB} \right)^{ab} (S_2^\mathrm{S} )_{AB} \\ 
  &{} + \gamma_{34}\,\left( C_2^\mathrm{S} T^{AB} \right)^{ab} (C_2^\mathrm{G} )_{AB} +   \gamma_{35}\,\left( C_2^\mathrm{S} C_2^\mathrm{S} C_2^\mathrm{S} \right)^{ab} \\
  &{} +\gamma_{36}\, (T^{AB})^{ab} \, \tr\left(t_{BA} C_2^\mathrm{F}\right) +\gamma_{37}\, (T^{AB})^{ab} \, \tr\left(T_{BA} C_2^\mathrm{S}\right)\,.
\end{aligned}\label{eq:ad-g6l}
\end{align}
The coefficient $\gamma_7$ appears twice to ensure that $\hat{\gamma}_\phi$ is symmetric (Hermitian) consistent with our choice for fixing the ambiguity. In fact, both tensor structures correspond to the first diagram in \eq{asymmetric-legs}. With other choices, the antisymmetric combination would also appear in the scalar leg corrections as discussed in \Sec{beta-ambiguity}.

The vertex corrections are grouped into several terms according to the coupling content:
      \begin{equation}\label{eq:vertex-general}
        \begin{aligned}
          \tilde{\beta}_\lambda^{(3) abcd} = &{}\;\;\tilde{\beta}_\lambda^{(3) abcd} \big|_{g=0} +  \tilde{\beta}_{g^2 y^6}^{abcd} + \tilde{\beta}_{g^2 y^4 \lambda}^{abcd} + \tilde{\beta}_{g^2 y^2 \lambda^2}^{abcd} + \tilde{\beta}_{g^2 \lambda^3}^{abcd} \\
          &{} +  \tilde{\beta}_{g^4 y^4 }^{abcd} +  \tilde{\beta}_{g^4 y^2 \lambda }^{abcd} + \tilde{\beta}_{g^4 \lambda^2 }^{abcd} +  \tilde{\beta}_{g^6 y^2 }^{abcd} + \tilde{\beta}_{g^6 \lambda }^{abcd} +\tilde{\beta}_{g^8 }^{abcd} .
        \end{aligned}
      \end{equation}
The basis for the gaugeless parts are found in \cite{Steudtner:2021fzs,Jack:2023zjt}. In our basis, the remaining terms are parametrised as 

      The operator $\mathcal{S}_n$ refers to summing over all $n$ inequivalent permutations of the indices $a$, $b$, $c$ and $d$, e.g.,
\begin{equation}
  \mathcal{S}_3\, \lambda_{abef}\lambda_{cdef} = \lambda_{abef}\lambda_{cdef} + \lambda_{acef}\lambda_{bdef} + \lambda_{adef}\lambda_{bcef}\,.
\end{equation}

      By comparing to explicit loop computations, the coefficients in~\eq{ad-general} and~\eq{vertex-general} can be identified. In the present case, we have used the $\overline{\text{MS}}$ renormalisation scheme with the semi-na\"ive $ \gamma_5 $ scheme and the Hermitianity condition to fix the RG ambiguity, and the listed coefficients refer exclusively to this case.
      Starting with the leg corrections, the gaugeless coefficients are listed in~\cite{Steudtner:2021fzs,Jack:2023zjt}, where they can be extracted in full from, e.g.,~\cite{Herren:2017uxn} and cross-checked with our other results. The remaining coefficients in~(\ref{eq:ad-g2y4}--\ref{eq:ad-g6l}), extracted from the toy-model comparison, are found to be   
      \begin{align}\label{eq:leg_coeff}
        \gamma_{1\phantom{0}} &= -\tfrac34\,, 
        &\gamma_{2\phantom{0}} &= -\tfrac98 +3\zeta_3\,, 
        &\gamma_{3\phantom{0}}& = \tfrac58 -3\zeta_3\,, 
        \notag \\
        \gamma_{4\phantom{0}}& = \tfrac{41}8 -6\zeta_3\,,
        &\gamma_{5\phantom{0}} &= \tfrac54+3\zeta_3\,, 
        &\gamma_{6\phantom{0}}& = -\tfrac{83}8 +3\zeta_3\,, 
        \notag \\ 
        \gamma_{7\phantom{0}}& = -\tfrac{13}8\,,
        &\gamma_{8\phantom{0}}& = \tfrac{11}2 
        &\gamma_{9\phantom{0}} &= -5 \,,
        \notag \\ \gamma_{10}&= -4 \,, 
        &\gamma_{11}& = \tfrac{19}{16}\,, 
        &\gamma_{12}&= -\tfrac{17}{48} \,,
        \notag \\
        \gamma_{13} &= \tfrac18\,, 
        &\gamma_{14}& = \tfrac3{32}\,, 
        &\gamma_{15}& = -\tfrac{59}{32} + \tfrac92 \zeta_3\,, 
        \notag \\
        \gamma_{16}& = -2\,,
        &\gamma_{17} &= -\tfrac{11}8\,, 
        &\gamma_{18}& = \tfrac{77}4 - 9 \zeta_3\,, 
        \notag \\
        \gamma_{19}& = \tfrac{123}{16} - \tfrac{15}2\zeta_3\,, \qquad
        &\gamma_{20}& = - \tfrac{29}2 + 3 \zeta_3\,, 
        &\gamma_{21}& = - \tfrac38 + 15 \zeta_3\,, 
        \notag \\
        \gamma_{22} &= - \tfrac{11}2\,, 
        &\gamma_{23}& = - \tfrac98\,,
        &\gamma_{24}& = - \tfrac{23}8\,, 
        \notag \\
        \gamma_{25}& = \tfrac{13}4 - 5 \zeta_3\,, 
        &\gamma_{26}& = \tfrac{35}{27}\,,
        &\gamma_{27}& = \tfrac{113}{108} + 24 \zeta_3\,, 
        \notag \\
        \gamma_{28} &= \tfrac{19}{27} \,,
        &\gamma_{29}& = - \tfrac{6059}{216} - 12 \zeta_3\,,\qquad
        &\gamma_{30}& = -\tfrac{91}{432} + 6 \zeta_3\,, 
        \notag \\
        \gamma_{31} &= \tfrac{5}{432} \,,
        &\gamma_{32}&= - \tfrac72\,,
        &\gamma_{33}& = - \tfrac{11}8\,,
        \notag \\
        \gamma_{34} &= -\tfrac{39}2 - 6 \zeta_3\,,
        &\gamma_{35}& = \tfrac32 + 3 \zeta_3\,,
        &\gamma_{36}& = \tfrac{45}4 - 12\zeta_3\,, 
        \notag \\
        \gamma_{37} &= \tfrac{33}2 - 6 \zeta_3 \,. 
      \end{align}
As for the vertex corrections, we find agreement with the the gaugeless coefficients presented in~\cite{Jack:2023zjt}. We have confirmed these results by computing the vertex correction directly in our gaugeless model. This confirms the gaugeless limit of the computations \cite{Jack:1990eb,Chetyrkin:2012rz,Bednyakov:2013eba,Chetyrkin:2013wya,Bednyakov:2013cpa,Bednyakov:2014pia,Zerf:2017zqi,Mihaila:2017ble,Jack:1996qq,Parkes:1985hh} and implies that $\mathcal{N}=1$ and $\mathcal{N}=\tfrac12$ supersymmetry relations~\cite{Fei:2016sgs,Jack:2023zjt} hold at three-loop order in the absence of gauge interactions.
The remaining vertex are again extracted from the toy-model \befs and cross-checked with the SM result~\cite{Bednyakov:2013cpa}. They are 
      \begin{align}\label{eq:vertex_coeff}
        \beta_{1\phantom{00}} &= 0\,, 
        &\beta_{2\phantom{00}} &= -7 + 6 \zeta_3\,,
        &\beta_{3\phantom{00}} &= -\tfrac{85}8 + 12 \zeta_3\,,
        \notag\\\beta_{4\phantom{00}} &= 4\,,
        &\beta_{5\phantom{00}} &= \tfrac{17}4 - 12 \zeta_3\,,
        &\beta_{6\phantom{00}} &= -8 \,,
        \notag\\\beta_{7\phantom{00}} &= \tfrac{65}2 - 12 \zeta_3\,,
        &\beta_{8\phantom{00}} &= 3 - 6 \zeta_3 \,,
        &\beta_{9\phantom{00}} &= -18 - 12 \zeta_3 \,,
        \notag\\\beta_{10\phantom{0}} &= 2 - 12 \zeta_3\,,
        &\beta_{11\phantom{0}} &= -6 + 12 \zeta_3\,,
        &\beta_{12\phantom{0}} &= -7 + 12 \zeta_3\,,
        \notag\\\beta_{13\phantom{0}} &= -40 + 24 \zeta_3\,,
        &\beta_{14\phantom{0}} &= 23 - 12 \zeta_3\,,
        &\beta_{15\phantom{0}} &= 10 +12 \zeta_3\,,
        \notag\\\beta_{16\phantom{0}} &=  -1 + 6 \zeta_3\,,
        &\beta_{17\phantom{0}} &= 8\,,
        &\beta_{18\phantom{0}} &= 8 - 24\zeta_3\,,
        \notag\\\beta_{19\phantom{0}} &=  22 - 24 \zeta_3\,,
        &\beta_{20\phantom{0}} &= 0\,,
        &\beta_{21\phantom{0}} &= 3\,,
        \notag\\\beta_{22\phantom{0}} &=  0\,,
        &\beta_{23\phantom{0}} &= 36 \zeta_3\,,
        &\beta_{24\phantom{0}} &= 8 + 12 \zeta_3\,,
        \notag\\\beta_{25\phantom{0}} &=  18 - 12 \zeta_3\,,
        &\beta_{26\phantom{0}} &= -8 + 24 \zeta_3\,,
        &\beta_{27\phantom{0}} &= 36 \zeta_3\,,
        \notag\\\beta_{28\phantom{0}} &=  36 \zeta_3\,,
        &\beta_{29\phantom{0}} &= - 36 \zeta_3\,,
        &\beta_{30\phantom{0}} &= 6 - 12 \zeta_3\,,
        \notag\\\beta_{31\phantom{0}} &=  22 - 36 \zeta_3\,,
        &\beta_{32\phantom{0}} &= - 6 - 12 \zeta_3\,,
        &\beta_{33\phantom{0}} &= -3 + 6 \zeta_3\,,
        \notag\\\beta_{34\phantom{0}} &=  0\,,
        &\beta_{35\phantom{0}} &= -21 + 30 \zeta_3\,,
        &\beta_{36\phantom{0}} &= 0 \,,
        \notag\\\beta_{37\phantom{0}} &=  -12 + 24 \zeta_3\,,
        &\beta_{38\phantom{0}} &= 4 + 24 \zeta_3\,,
        &\beta_{39\phantom{0}} &= 18 - 6 \zeta_3\,,
        \notag\\\beta_{40\phantom{0}} &=  -6 - 6 \zeta_3\,,
        &\beta_{41\phantom{0}} &= 0\,,
        &\beta_{42\phantom{0}} &= -26 + 30 \zeta_3\,,
        \notag\\\beta_{43\phantom{0}} &=  -16 - 12 \zeta_3\,,
        &\beta_{44\phantom{0}} &= 0 \,,
        &\beta_{45\phantom{0}} &= 0 \,,
        \notag\\\beta_{46\phantom{0}} &=  4\,,
        &\beta_{47\phantom{0}} &= \tfrac52 - 9 \zeta_3 \,,
        &\beta_{48\phantom{0}} &= - \tfrac{51}4 + 12 \zeta_3\,,
        \notag\\\beta_{49\phantom{0}} &=  0\,,
        &\beta_{50\phantom{0}} &= \tfrac12 + 6 \zeta_3\,,
        &\beta_{51\phantom{0}} &= 0\,,
        \notag\\\beta_{52\phantom{0}} &=  0\,,
        &\beta_{53\phantom{0}} &= \tfrac{11}4 - 3 \zeta_3\,,
        &\beta_{54\phantom{0}} &= 0 \,,
        \notag\\\beta_{55\phantom{0}} &=  \tfrac{25}4 - 9 \zeta_3\,,
        &\beta_{56\phantom{0}} &= -17 + 12 \zeta_3\,,
        &\beta_{57\phantom{0}} &= 0 \,,
        \notag\\\beta_{58\phantom{0}} &=  \tfrac52\,,
        &\beta_{59\phantom{0}} &= \tfrac{15}2 - 12 \zeta_3\,,
        &\beta_{60\phantom{0}} &= 9\,,
        \notag\\\beta_{61\phantom{0}} &=  \tfrac{19}4\,,
        &\beta_{62\phantom{0}} &= -\tfrac{171}2 + 48 \zeta_3\,,
        &\beta_{63\phantom{0}} &= -6\,,
        \notag\\\beta_{64\phantom{0}} &=  -3\,,
        &\beta_{65\phantom{0}} &= 56 - 36 \zeta_3\,,
        &\beta_{66\phantom{0}} &= -8\,,
        \notag\\\beta_{67\phantom{0}} &=  -2\,,
        &\beta_{68\phantom{0}} &= 62 - 48 \zeta_3\,,
        &\beta_{69\phantom{0}} &= \tfrac{89}2 - 9 \zeta_3\,,
        \notag\\\beta_{70\phantom{0}} &=  90 - 138 \zeta_3\,,
        &\beta_{71\phantom{0}} &= 94 - 162 \zeta_3\,,
        &\beta_{72\phantom{0}} &= - 59 + 99 \zeta_3\,,
        \notag\\\beta_{73\phantom{0}} &=  - 41 + 51 \zeta_3\,, \qquad 
        &\beta_{74\phantom{0}} &= 37 - 57 \zeta_3\,,
        &\beta_{75\phantom{0}} &= 34 - 60 \zeta_3\,,
        \notag\\\beta_{76\phantom{0}} &=  -19 + 6 \zeta_3\,,
        &\beta_{77\phantom{0}} &= -72 + 102 \zeta_3\,, \qquad 
        &\beta_{78\phantom{0}} &= -42 18 \zeta_3\,,
        \notag\\\beta_{79\phantom{0}} &=  43 - 69 \zeta_3\,,
        &\beta_{80\phantom{0}} &= -\tfrac14\,,
        &\beta_{81\phantom{0}} &= -\tfrac52\,,
        \notag\\\beta_{82\phantom{0}} &=  24 \,,
        &\beta_{83\phantom{0}} &= 0 \,,
        &\beta_{84\phantom{0}} &= \tfrac32 + 12 \zeta_3\,,
        \notag\\\beta_{85\phantom{0}} &=  11 \,,
        &\beta_{86\phantom{0}} &= -\tfrac{131}{16}\,,
        &\beta_{87\phantom{0}} &= 44 - 12 \zeta_3\,,
        \notag\\\beta_{88\phantom{0}} &=  24\,,
        &\beta_{89\phantom{0}} &= -\tfrac{39}8 - 6 \zeta_3\,,
        &\beta_{90\phantom{0}} &= 18  -24 \zeta_3\,,
        \notag\\\beta_{91\phantom{0}} &=  140 - 84 \zeta_3\,,
        &\beta_{92\phantom{0}} &= 0\,,
        &\beta_{93\phantom{0}} &= -42 + 90 \zeta_3\,,
        \notag\\\beta_{94\phantom{0}} &=  -52 + 48 \zeta_3\,,
        &\beta_{95\phantom{0}} &= -4 + 12 \zeta_3\,,
        &\beta_{96\phantom{0}} &= 0\,,
        \notag\\\beta_{97\phantom{0}} &=  0\,,
        &\beta_{98\phantom{0}} &= -\tfrac12\,,
        &\beta_{99\phantom{0}} &= -10\,,
        \notag\\\beta_{100} &=  -18 + 12 \zeta_3\,,
        &\beta_{101} &= -\tfrac{19}2\,,
        &\beta_{102} &= -10\,,
        \notag\\\beta_{103} &=  44 - 48 \zeta_3\,,
        &\beta_{104} &= \tfrac12 \,,
        &\beta_{105} &= \tfrac38\,,
        \notag\\\beta_{106} &=  - \tfrac{41}8 + \tfrac92 \zeta_3\,,
        &\beta_{107} &= - \tfrac{17}2\,,
        &\beta_{108} &= - \tfrac{35}8\,,
        \notag\\\beta_{109} &=  \tfrac{483}8 - 24 \zeta_3\,,
        &\beta_{110} &= 1\,,
        &\beta_{111} &= \tfrac34\,,
        \notag\\\beta_{112} &=  -\tfrac{41}4 + 9 \zeta_3\,,
        &\beta_{113} &= \tfrac{55}8 + \tfrac92 \zeta_3\,,
        &\beta_{114} &= \tfrac{23}4 - 3 \zeta_3\,,
        \notag\\\beta_{115} &=  0\,,
        &\beta_{116} &= 0\,,
        &\beta_{117} &= -\tfrac{57}4 + 42 \zeta_3\,,
        \notag\\\beta_{118} &=  -\tfrac{45}2 + 6 \zeta_3\,,
        &\beta_{119} &= -\tfrac{39}2 - 6 \zeta_3\,,
        &\beta_{120} &= - 6 - 12 \zeta_3\,,
        \notag\\\beta_{121} &=  -1\,,
        &\beta_{122} &= 0\,,
        &\beta_{123} &= -13 + 3 \zeta_3\,,
        \notag\\\beta_{124} &=  \tfrac{43}{48} - \tfrac12 \zeta_3\,,
        &\beta_{125} &= \tfrac{39}2 + 6 \zeta_3\,,
        &\beta_{126} &= - \tfrac{21}2\,,
        \notag\\\beta_{127} &=  -3 - 48 \zeta_3\,,
        &\beta_{128} &= -16 + 6 \zeta_3\,,
        &\beta_{129} &= 1 \,,
        \notag\\\beta_{130} &=  0\,,
        &\beta_{131} &= \tfrac{87}8 + \tfrac{45}2 \zeta_3\,,
        &\beta_{132} &= 4\,,
        \notag\\\beta_{133} &=  3\,,
        &\beta_{134} &= -82 - 12 \zeta_3\,,
        &\beta_{135} &= -\tfrac72 + 15 \zeta_3\,,
        \notag\\\beta_{136} &=  - \tfrac{83}4 - 21 \zeta_3\,,
        &\beta_{137} &= \tfrac{241}8 - 18 \zeta_3\,,
        &\beta_{138} &= 9\,,
        \notag\\\beta_{139} &=  -30\,,
        &\beta_{140} &= 19 - 12 \zeta_3\,,
        &\beta_{141} &= 83 + 12 \zeta_3\,,
        \notag\\\beta_{142} &=  \tfrac{89}4 + 21 \zeta_3\,,
        &\beta_{143} &= -136 - 24 \zeta_3\,,
        &\beta_{144} &= \tfrac94\,,
        \notag\\\beta_{145} &=  \tfrac{29}4\,,
        &\beta_{146} &= 0\,,
        &\beta_{147} &= 0\,,
        \notag\\\beta_{148} &=  0\,,
        &\beta_{149} &= 4\,,
        &\beta_{150} &= 0\,,
        \notag\\\beta_{151} &=  \tfrac{95}2 + 60 \zeta_3\,,
        &\beta_{152} &= 20\,,
        &\beta_{153} &= 9\,,
        \notag\\\beta_{154} &=  -224 + 192 \zeta_3\,,
        &\beta_{155} &= 3 - 27 \zeta_3\,,
        &\beta_{156} &= -6 + 54 \zeta_3\,,
        \notag\\\beta_{157} &=  0\,,
        &\beta_{158} &= - \tfrac{165}2\,,
        &\beta_{159} &= 0\,,
        \notag\\\beta_{160} &=  76 - 36 \zeta_3\,,
        &\beta_{161} &= 155 - 12 \zeta_3\,,
        &\beta_{162} &= 171\,,
        \notag\\\beta_{163} &=  44 - 192 \zeta_3\,,
        &\beta_{164} &= -79 + 96 \zeta_3\,,
        &\beta_{165} &= 8 + 12 \zeta_3\,,
        \notag\\\beta_{166} &=  - \tfrac{40}{27}\,,
        &\beta_{167} &= - \tfrac{2038}{27} - 96 \zeta_3\,,
        &\beta_{168} &= - \tfrac{14}{27}\,,
        \notag\\\beta_{169} &=  \tfrac{20213}{27} - \tfrac{297}2 \zeta_3\,,  \qquad 
        &\beta_{170} &= - \tfrac{1499}{54} - \tfrac{45}2 \zeta_3\,,  \qquad 
        &\beta_{171} &= \tfrac7{54}\,,
        \notag\\\beta_{172} &=  - \tfrac{40}{27}\,,
        &\beta_{173} &= - \tfrac7{27}\,,
        &\beta_{174} &= \tfrac{581}{54}\,,
        \notag\\\beta_{175} &=  \tfrac{539}{216}\,,
        &\beta_{176} &= \tfrac7{54}\,,
        &\beta_{177} &= - \tfrac{35923}{108} + \tfrac{399}2 \zeta_3\,,
        \notag\\\beta_{178} &=  \tfrac{157}4 + 36 \zeta_3\,,
        &\beta_{179} &= \tfrac{327}{16} + \tfrac{27}4 \zeta_3\,,
        &\beta_{180} &= - \tfrac{7587}{16} + \tfrac{219}2 \zeta_3\,,
        \notag\\\beta_{181} &=  -6\,,
        &\beta_{182} &= 0 \,,
        &\beta_{183} &= - \tfrac{305}4 - 30 \zeta_3\,,
        \notag\\\beta_{184} &=  - 51 - 72 \zeta_3\,,
        &\beta_{185} &= - \tfrac{105}4 - \tfrac{27}2 \zeta_3\,,
        &\beta_{186} &= \tfrac{3051}4 - 117 \zeta_3\,,
        \notag\\\beta_{187} &=  \tfrac{3225}4 - 294 \zeta_3\,,
        &\beta_{188} &= -242 + 93 \zeta_3\,,
        &\beta_{189} &= - \tfrac{855}2 + 153 \zeta_3\,,
        \notag\\\beta_{190} &=  - \tfrac{51}2 + 24 \zeta_3\,,
        &\beta_{191} &= -39 + 12 \zeta_3\,,
        &\beta_{192} &= -32 + 24 \zeta_3\,,
        \notag\\\beta_{193} &=  -1 + 15 \zeta_3\,,
        &\beta_{194} &= 466 - 186 \zeta_3\,,
        &\beta_{195} &= -68 + 156 \zeta_3\,.
      \end{align}

Using the dummy-field technique~\cite{Martin:1993zk,Luo:2002ti,Schienbein:2018fsw}, scalar mass, cubic, and tadpole coupling \befs follow immediately from $\beta_\lambda$. 
For example, consider the scalar cubic coupling, which we define promoting one scalar field component $\phi_{a=0}$ to be non-dynamic and constant: $h^{abc} = \lambda^{abc0}\,\phi_0$.
There is no leg correction to such a constant (dummy) field, but otherwise the definition~\eq{def-beta-lambda} still applies
\begin{equation} \label{eq:trilinear_beta}
  \beta_h^{abc} = \tilde{\beta}_\lambda^{abc0}  + \hat{\gamma}_\phi^{ae} h^{ebc} + \hat{\gamma}_\phi^{be} h^{eac} + \hat{\gamma}_\phi^{ce} h^{eab} \,,
\end{equation}
where $\tilde{\beta}_\lambda^{abc0}$ are the quartic vertex correction with the dummy field inserted as external leg. Importantly, this observation relies on the dummy field being neutral, which ensures that the leg corrections to the dummy field $\gamma_\phi^{a0}$ are indentical to the leg corrections  $ \hat{\gamma}_\phi^{a0} $ in $\beta_\lambda$:  
\begin{equation}\label{eq:uncharged-legs}
  \hat{\gamma}_\phi^{a0} = \gamma_\phi^{a0}. 
\end{equation}
This is because both legs in $\gamma_\phi^{a0}$ are neutral by gauge invariance such that gauge relations~\eq{gauge-relations} cannot modify $\gamma_\phi$: gauge lines contained in vertex corrections cannot cross over a neutral scalar line via the transformations \eq{gauge-relations} to become leg corrections or vice versa, 
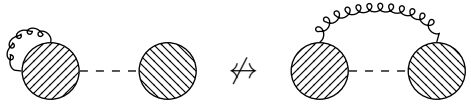
\newline
\begin{center}
  \begin{tabular}{lcr}
    \begin{tikzpicture}[baseline=-0.25em]
      \begin{feynhand}
        \vertex [NEblob] (O1) at (0em,0em) {};
        \vertex [NWblob] (O1) at (4em,0em) {};
        \vertex [dot] (v1) at (0em,1em) ;
        \vertex [dot] (v2) at (-1em,0em);
        \vertex [dot] (c1) at (1em,0em);
        \vertex [dot] (c2) at (3em,0em);
        \propag [gluon] (v1) to [half right, looseness=2] (v2);
        \propag [scalar] (c1) to (c2);
      \end{feynhand}
    \end{tikzpicture} & 
    $\not \leftrightarrow$ &
    \begin{tikzpicture}[baseline=-0.25em]
      \begin{feynhand}
        \vertex [NEblob] (O1) at (0em,0em) {};
        \vertex [NWblob] (O1) at (4em,0em) {};
        \vertex [dot] (v1) at (0em,1em) ;
        \vertex [dot] (v2) at (4em,1em);
        \vertex [dot] (c1) at (1em,0em);
        \vertex [dot] (c2) at (3em,0em);
        \propag [gluon] (v1) to [half left, looseness=1] (v2);
        \propag [scalar] (c1) to (c2);
      \end{feynhand}
    \end{tikzpicture}\,.
  \end{tabular}
\end{center}
 Hence, \eqref{eq:trilinear_beta} correctly subtracts all leg correction of the dummy field.

\section{Outlook/Conclusion}\label{sec:Outlook}

The main results of this work are the 3-loop quartic template \befs specified in Eqs.~\eq{def-beta-lambda}--\eq{vertex_coeff}, valid for any renormalisable, four-dimensional theory. For convenience, this result is also translated into the notation of Ref.~\cite{Poole:2019kcm,Davies:2021mnc} in \App{NotationAnders}. 
Due to its advanced complexity, the result is not particularly useful for human consumption. Rather, its true potential unfolds when implemented in a software package automating the generation of coupling tensors and index contractions.
To this end, the three-loop quartic \bef have been made available as part of the toolkits \texttt{RGBeta} (\texttt{v1.2.0})~\cite{Thomsen:2021ncy} and \texttt{FoRGEr}~\cite{Steudtner:FoRGEr}, which serve as reference implementations.
With this result, the \befs of the marginal couplings in renormalisable four-dimensional theories are now fully known up to four-loop orders for gauge couplings and three-loop orders for both Yukawa and quartic couplings (see also \cite{Bednyakov:2021qxa,Davies:2021mnc}). 

On the other hand, general RGEs for scalar and fermion anomalous dimensions~\cite{Machacek:1983tz} as well as vacuum-expectation-values~(VEVs)~\cite{Sperling:2013eva,Sperling:2013xqa} are only available up to two loops. Generic results for gauge-fixing parameters are absent entirely in the literature. While these quantities are manifest gauge-dependent, they are by no means less important. For instance, anomalous dimensions may turn physical at critical points, and are required to leverage the RG-invariance of effective potentials.
The RG evolution of VEVs and gauge parameters is generally intertwined. More so,
RGEs for gauge-fixing parameters are required for the running of any gauge-dependent quantity. Overall, the determination of template RGEs for anomalous dimensions, VEVs and gauge parameters to three-loop order is an alley worth pursuing in future works.

The utilisation of automated software removes the complication of computing \befs by hand. This ushers in a new age of complete three-loop precision when investigating the RG flow of renormalisable QFTs. In fact, the same strategy of computing template expressions and building software around them to achieve multi-loop precision has been fruitful in many areas. A notable complement to this work are the efforts to automate matching between renormalisable and/or non-renormalisable QFTs through packages such as \texttt{Matchmakereft}~\cite{Carmona:2021xtq} and \texttt{Matchete}~\cite{Fuentes-Martin:2022jrf}.

Generic higher loop corrections to $\beta$-functions will have deal with the $\gamma_5$-problem. While five-loop gauge- and four-loop Yukawa $\beta$-functions are in principle related to the results obtained here by Weyl-consistency relations~\cite{Osborn:1989td,Jack:1990eb,Osborn:1991gm}, it is not clear if all $\gamma_5$-ambiguities can be eliminated as was the case at lower loops~\cite{Poole:2019txl}.
Alternatively, the Breitenlohner--Maison--'t Hooft--Veltman treatment of $\gamma_5$~\cite{tHooft:1972tcz,Breitenlohner:1977hr} might have to be employed. A recent example of such a computation is found in~\cite{Stockinger:2023ndm}.

\section*{Acknowledgments}
TS in indebted to Dominik Hellmann and Mustafa Tabet for working overtime to minimise disruptions of the computations in this project. 
AET is funded by the Swiss National Science Foundation (SNSF) through the Ambizione grant ``Matching and Running: Improved Precision in the Hunt for New Physics,'' project number 209042.

\appendix
\section{3-loop Quartic Beta Function}\label{sec:NotationAnders}
Here we report the explicit tensor structure parametrization of the 3-loop quartic \bef following the notation of~\cite{Poole:2019kcm,Davies:2021mnc}, which is how it is implemented in \texttt{RGBeta}~\cite{Thomsen:2021ncy}. This setup is subtly different from the notation, we employed in Sec.~\ref{sec:Formalism} of the present paper. And the results of this appendix is entirely redundant. Nevertheless, for the reader following the notation of these papers, this appendix will no doubt prove convenient. The notation employed in this appendix is that of Ref.~\cite{Davies:2021mnc}.

\subsection{Parametrization of the Quartic \texorpdfstring{$ \boldsymbol{\beta} $-Function}{Beta Function}}
The code \texttt{GRAFER}~\cite{Poole:2019kcm} generates a basis of tensor structures for the quartic \bef.
This basis is entirely general and do not make any assumption about how the RG ambiguity is fixed. All tensors are symmetrized over all permutations of the four open indices. We have  

{\small


\subsection{Coefficients of the Quartic \texorpdfstring{$ \boldsymbol{\beta} $-Function}{Beta Function}}
The coefficients of the \texttt{GRAFER} basis quartic \bef are given below. These coefficients reflect the particular choice of using Hermitian scalar wave-function renormalisation factors to fix the RG ambiguity:

\begin{align*}
\cofq{3}{1} & = 936 \zeta_3 - 408 & 
\cofq{3}{2} & = 2796 - 1116 \zeta_3 &
\cofq{3}{3} & = 180 \zeta_3 - 12 &
\cofq{3}{4} & = 288 \zeta_3 - 384 \\
\cofq{3}{5} & = 144 \zeta_3 - 468 &
\cofq{3}{6} & = 288 \zeta_3 - 306 &
\cofq{3}{7} & = 3672 \zeta_3 - 10260 &
\cofq{3}{8} & = 1116 \zeta_3 - 2904 \\
\cofq{3}{9} & = 9675 - 3528 \zeta_3 &
\cofq{3}{10} & = 9153 - 1404 \zeta_3 &
\cofq{3}{11} & = -315 - 162 \zeta_3 &
\cofq{3}{12} & = -306 - 432 \zeta_3 \\
\cofq{3}{13} & = -1830 - 720 \zeta_3 &
\cofq{3}{14} & = 2628 \zeta_3 - \dfrac{22761}{2} &
\cofq{3}{15} & = 0 &
\cofq{3}{16} & = \dfrac{981}{2} + 162 \zeta_3 \\
\cofq{3}{17} & = -72 &
\cofq{3}{18} & = 471 + 432 \zeta_3 &
\cofq{3}{19} & = 1197 \zeta_3 - \dfrac{35923}{18} &
\cofq{3}{20} & = \dfrac{7}{9} \\
\cofq{3}{21} & = \dfrac{14}{9} &
\cofq{3}{22} & = -\dfrac{2998}{9} - 270 \zeta_3 &
\cofq{3}{23} & = \dfrac{80852}{9} - 1782 \zeta_3 &
\cofq{3}{24} & = \dfrac{539}{18} \\
\cofq{3}{25} & = -\dfrac{28}{9} &
\cofq{3}{26} & = -\dfrac{4076}{9} - 576 \zeta_3 &
\cofq{3}{27} & = \dfrac{581}{9} &
\cofq{3}{28} & = -\dfrac{14}{9} \\
\cofq{3}{29} & = -\dfrac{20}{9} &
\cofq{3}{30} & = -\dfrac{40}{9} &
\cofq{3}{31} & = 192 + 288 \zeta_3 &
\cofq{3}{32} & = 1152 \zeta_3 - 948 \\
\cofq{3}{33} & = 264 - 1152 \zeta_3 &
\cofq{3}{34} & = 1026 &
\cofq{3}{35} & = 1860 - 144 \zeta_3 &
\cofq{3}{36} & = 912 - 432 \zeta_3 \\
\cofq{3}{37} & = 0 &
\cofq{3}{38} & = 66 - 24 \zeta_3 &
\cofq{3}{39} & = -990 &
\cofq{3}{40} & = 45 - 48 \zeta_3 \\
\cofq{3}{41} & = 0 &
\cofq{3}{42} & = 216 \zeta_3 - 24 &
\cofq{3}{43} & = 36 - 324 \zeta_3 &
\cofq{3}{44} & = 6 + 12 \zeta_3 \\
\cofq{3}{45} & = 1152 \zeta_3 - 1344 &
\cofq{3}{46} & = 54 &
\cofq{3}{47} & = 60 &
\cofq{3}{48} & = 285 + 360 \zeta_3 \\
\cofq{3}{49} & = 0 &
\cofq{3}{50} & = 12 &
\cofq{3}{51} & = 0 &
\cofq{3}{52} & = 0 \\
\cofq{3}{53} & = -78 - 24 \zeta_3 &
\cofq{3}{54} & = -\dfrac{11}{2} &
\cofq{3}{55} & = 0 &
\cofq{3}{56} & = -7 \\
\cofq{3}{57} & = \dfrac{5}{108} &
\cofq{3}{58} & = 24 \zeta_3 - \dfrac{91}{108} &
\cofq{3}{59} & = -\dfrac{6059}{54} - 48 \zeta_3 &
\cofq{3}{60} & = \dfrac{38}{27} \\
\cofq{3}{61} & = \dfrac{113}{54} + 48 \zeta_3 &
\cofq{3}{62} & = \dfrac{35}{27} &
\cofq{3}{63} & = 72 \zeta_3 - 192 &
\cofq{3}{64} & = -36 - 576 \zeta_3 \\
\cofq{3}{65} & = -126 &
\cofq{3}{66} & = 117 + 36 \zeta_3 &
\cofq{3}{67} & = 13 - 24 \zeta_3 &
\cofq{3}{68} & = \dfrac{43}{2} - 12 \zeta_3 \\
\cofq{3}{69} & = 36 \zeta_3 - 156 &
\cofq{3}{70} & = 0 &
\cofq{3}{71} & = -3 &
\cofq{3}{72} & = -72 - 144 \zeta_3 \\
\cofq{3}{73} & = -234 - 72 \zeta_3 &
\cofq{3}{74} & = 18 \zeta_3 - \dfrac{135}{2} &
\cofq{3}{75} & = 126 \zeta_3 - \dfrac{171}{4} &
\cofq{3}{76} & = 0 \\
\cofq{3}{77} & = 0 &
\cofq{3}{78} & = \dfrac{69}{2} - 18 \zeta_3 &
\cofq{3}{79} & = \dfrac{165}{2} + 54 \zeta_3 &
\cofq{3}{80} & = 54 \zeta_3 - \dfrac{123}{2} \\
\cofq{3}{81} & = \dfrac{9}{2} &
\cofq{3}{82} & = 3 &
\cofq{3}{83} & = \dfrac{1449}{8} - 72 \zeta_3 &
\cofq{3}{84} & = -\dfrac{105}{8} \\
\cofq{3}{85} & = -\dfrac{51}{4} &
\cofq{3}{86} & = 54 \zeta_3 - \dfrac{123}{2} &
\cofq{3}{87} & = \dfrac{9}{2} &
\cofq{3}{88} & = 3 \\
\cofq{3}{89} & = 90 - 144 \zeta_3 &
\cofq{3}{90} & = 60 &
\cofq{3}{91} & = 0 &
\cofq{3}{92} & = 72 \zeta_3 - 102 \\
\cofq{3}{93} & = -\dfrac{17}{12} &
\cofq{3}{94} & = \dfrac{19}{4} &
\cofq{3}{95} & = 75 - 108 \zeta_3 &
\cofq{3}{96} & = 0 \\
\cofq{3}{97} & = 33 - 36 \zeta_3 &
\cofq{3}{98} & = 0 &
\cofq{3}{99} & = 0 &
\cofq{3}{100} & = 12 \zeta_3 \\
\cofq{3}{101} & = -\dfrac{9}{8} &
\cofq{3}{102} & = \dfrac{3}{2} &
\cofq{3}{103} & = 24 &
\cofq{3}{104} & = -\dfrac{1}{4} \\
\cofq{3}{105} & = -3 &
\cofq{3}{106} & = -3 &
\cofq{3}{107} & = 0 &
\cofq{3}{108} & = 0 \\
\cofq{3}{109} & = 87 &
\cofq{3}{110} & = 27 &
\cofq{3}{111} & = -1632 - 288 \zeta_3 &
\cofq{3}{112} & = 534 + 504 \zeta_3 \\
\cofq{3}{113} & = 1992 + 288 \zeta_3 &
\cofq{3}{114} & = 456 - 288 \zeta_3 &
\cofq{3}{115} & = -360 &
\cofq{3}{116} & = 216 \\
\cofq{3}{117} & = 723 - 432 \zeta_3 &
\cofq{3}{118} & = -498 - 504 \zeta_3 &
\cofq{3}{119} & = 360 \zeta_3 - 84 &
\cofq{3}{120} & = 72 \\
\cofq{3}{121} & = -1968 - 288 \zeta_3 &
\cofq{3}{122} & = 48 &
\cofq{3}{123} & = 261 + 540 \zeta_3 &
\cofq{3}{124} & = 0 \\
\cofq{3}{125} & = 12 &
\cofq{3}{126} & = 528 - 576 \zeta_3 &
\cofq{3}{127} & = -60 &
\cofq{3}{128} & = -114 \\
\cofq{3}{129} & = 144 \zeta_3 - 216 &
\cofq{3}{130} & = -60 &
\cofq{3}{131} & = -\dfrac{23}{2} &
\cofq{3}{132} & = -\dfrac{9}{2} \\
\cofq{3}{133} & = -6 &
\cofq{3}{134} & = 0 &
\cofq{3}{135} & = -22 &
\cofq{3}{136} & = 60 \zeta_3 - \dfrac{3}{2} \\
\cofq{3}{137} & = 12 \zeta_3 - 58 &
\cofq{3}{138} & = \dfrac{123}{4} - 30 \zeta_3 &
\cofq{3}{139} & = 0 &
\cofq{3}{140} & = 77 - 36 \zeta_3 \\
\cofq{3}{141} & = -\dfrac{11}{2} &
\cofq{3}{142} & = -4 &
\cofq{3}{143} & = 18 \zeta_3 - \dfrac{59}{8} &
\cofq{3}{144} & = \dfrac{3}{8} \\
\cofq{3}{145} & = \dfrac{1}{4} &
\cofq{3}{146} & = 6 + 72 \zeta_3 &
\cofq{3}{147} & = 0 &
\cofq{3}{148} & = 36 \zeta_3 - \dfrac{153}{4} \\
\cofq{3}{149} & = 15 - 54 \zeta_3 &
\cofq{3}{150} & = 12 &
\cofq{3}{151} & = 0 &
\cofq{3}{152} & = 0 \\
\cofq{3}{153} & = 12 &
\cofq{3}{154} & = -\dfrac{5}{8} &
\cofq{3}{155} & = -6 &
\cofq{3}{156} & = 0 \\
\cofq{3}{157} & = 288 \zeta_3 - 96 &
\cofq{3}{158} & = 576 \zeta_3 - 624 &
\cofq{3}{159} & = 1080 \zeta_3 - 504 &
\cofq{3}{160} & = 0 \\
\cofq{3}{161} & = 840 - 504 \zeta_3 &
\cofq{3}{162} & = 216 - 288 \zeta_3 &
\cofq{3}{163} & = -117 - 144 \zeta_3 &
\cofq{3}{164} & = 288 \\
\cofq{3}{165} & = 1056 - 288 \zeta_3 &
\cofq{3}{166} & = -\dfrac{393}{2} &
\cofq{3}{167} & = 132 &
\cofq{3}{168} & = 36 + 288 \zeta_3 \\
\cofq{3}{169} & = 0 &
\cofq{3}{170} & = 144 &
\cofq{3}{171} & = -60 &
\cofq{3}{172} & = -3 \\
\cofq{3}{173} & = 1032 - 1656 \zeta_3 &
\cofq{3}{174} & = 216 \zeta_3 - 504 &
\cofq{3}{175} & = 2448 \zeta_3 - 1728 &
\cofq{3}{176} & = 72 \zeta_3 - 228 \\
\cofq{3}{177} & = 204 - 360 \zeta_3 &
\cofq{3}{178} & = 444 - 684 \zeta_3 &
\cofq{3}{179} & = 1224 \zeta_3 - 984 &
\cofq{3}{180} & = 2376 \zeta_3 - 1416 \\
\cofq{3}{181} & = 564 - 972 \zeta_3 &
\cofq{3}{182} & = 1080 - 1656 \zeta_3 &
\cofq{3}{183} & = 534 - 108 \zeta_3 &
\cofq{3}{184} & = 372 - 288 \zeta_3 \\
\cofq{3}{185} & = -12 &
\cofq{3}{186} & = -24 &
\cofq{3}{187} & = 672 - 432 \zeta_3 &
\cofq{3}{188} & = -36 \\
\cofq{3}{189} & = 576 \zeta_3 - 1026 &
\cofq{3}{190} & = 57 &
\cofq{3}{191} & = -36 &
\cofq{3}{192} & = 54 \\
\cofq{3}{193} & = -96 - 72 \zeta_3 &
\cofq{3}{194} & = 180 \zeta_3 - 156 &
\cofq{3}{195} & = 0 &
\cofq{3}{196} & = -72 - 72 \zeta_3 \\
\cofq{3}{197} & = 216 - 72 \zeta_3 &
\cofq{3}{198} & = 24 + 144 \zeta_3 &
\cofq{3}{199} & = 144 \zeta_3 - 72 &
\cofq{3}{200} & = -16 \\
\cofq{3}{201} & = 0 &
\cofq{3}{202} & = 360 \zeta_3 - 252 &
\cofq{3}{203} & = 0 &
\cofq{3}{204} & = -20 \\
\cofq{3}{205} & = 72 \zeta_3 - 36 &
\cofq{3}{206} & = 22 &
\cofq{3}{207} & = -36 - 72 \zeta_3 &
\cofq{3}{208} & = -\dfrac{13}{2} \\
\cofq{3}{209} & = 264 - 432 \zeta_3 &
\cofq{3}{210} & = 36 - 72 \zeta_3 &
\cofq{3}{211} & = -\dfrac{13}{2} &
\cofq{3}{212} & = 12 \zeta_3 - \dfrac{83}{2} \\
\cofq{3}{213} & = 5 + 12 \zeta_3 &
\cofq{3}{214} & = \dfrac{41}{2} - 24 \zeta_3 &
\cofq{3}{215} & = \dfrac{5}{2} - 12 \zeta_3 &
\cofq{3}{216} & = 12 \zeta_3 - \dfrac{9}{2} \\
\cofq{3}{217} & = -3 &
\cofq{3}{218} & = 9 \zeta_3 - 3 &
\cofq{3}{219} & = 72 \zeta_3 - 48 &
\cofq{3}{220} & = 12 \\
\cofq{3}{221} & = 0 &
\cofq{3}{222} & = 6 &
\cofq{3}{223} & = \dfrac{15}{4} &
\cofq{3}{224} & = \dfrac{5}{2} \\
\cofq{3}{225} & = 24 &
\cofq{3}{226} & = \dfrac{5}{2} &
\cofq{3}{227} & = \dfrac{75}{8} &
\cofq{3}{228} & = 36 \\
\cofq{3}{229} & = 0 &
\cofq{3}{230} & = -\dfrac{3}{4} &
\cofq{3}{231} & = -\dfrac{3}{2} &
\cofq{3}{232} & = -432 \zeta_3 \\
\cofq{3}{233} & = 864 \zeta_3 &
\cofq{3}{234} & = 864 \zeta_3 &
\cofq{3}{235} & = 288 \zeta_3 - 96 &
\cofq{3}{236} & = 432 - 288 \zeta_3 \\
\cofq{3}{237} & = 96 + 144 \zeta_3 &
\cofq{3}{238} & = 864 \zeta_3 &
\cofq{3}{239} & = 0 &
\cofq{3}{240} & = 72 \\
\cofq{3}{241} & = 0 &
\cofq{3}{242} & = 264 - 288 \zeta_3 &
\cofq{3}{243} & = 48 - 144 \zeta_3 &
\cofq{3}{244} & = 96 \\
\cofq{3}{245} & = 144 \zeta_3 - 24 &
\cofq{3}{246} & = 240 + 288 \zeta_3 &
\cofq{3}{247} & = 288 \zeta_3 - 480 &
\cofq{3}{248} & = 552 - 288 \zeta_3 \\
\cofq{3}{249} & = 288 \zeta_3 - 168 &
\cofq{3}{250} & = 288 \zeta_3 - 144 &
\cofq{3}{251} & = 24 - 144 \zeta_3 &
\cofq{3}{252} & = -432 - 288 \zeta_3 \\
\cofq{3}{253} & = 72 - 144 \zeta_3 &
\cofq{3}{254} & = 96 &
\cofq{3}{255} & = 390 - 144 \zeta_3 &
\cofq{3}{256} & = 288 \zeta_3 - 255 \\
\cofq{3}{257} & = -96 &
\cofq{3}{258} & = 144 \zeta_3 - 168 &
\cofq{3}{259} & = 51 - 144 \zeta_3 &
\cofq{3}{260} & = 0 \\
\cofq{3}{261} & = -96 \zeta_3 &
\cofq{3}{262} & = -288 \zeta_3 &
\cofq{3}{263} & = -48 &
\cofq{3}{264} & = 72 \zeta_3 - 48 \\
\cofq{3}{265} & = 72 \zeta_3 - 60 &
\cofq{3}{266} & = 6 \zeta_3 - 4 &
\cofq{3}{267} & = -48 &
\cofq{3}{268} & = 12 \\
\cofq{3}{269} & = -120 &
\cofq{3}{270} & = -60 &
\cofq{3}{271} & = -72 &
\cofq{3}{272} & = -24 \\
\cofq{3}{273} & = -12 &
\cofq{3}{274} & = -\dfrac{3}{2} &
\cofq{3}{275} & = -3 &
\cofq{3}{276} & = 7 \\
\cofq{3}{277} & = -6 &
\cofq{3}{278} & = -36 &
\cofq{3}{279} & = -18 &
\cofq{3}{280} & = \dfrac{7}{4} \\
\cofq{3}{281} & = \dfrac{5}{4} &
\cofq{3}{282} & = \dfrac{7}{4} &
\cofq{3}{283} & = 36 &
\cofq{3}{284} & = 4 \\
\cofq{3}{285} & = \dfrac{9}{4} &
\cofq{3}{286} & = \dfrac{1}{8} &
\cofq{3}{287} & = -\dfrac{3}{4} &
\cofq{3}{288} & = -12 \\
\cofq{3}{289} & = -72 &
\cofq{3}{290} & = -72 \zeta_3 &
\cofq{3}{291} & = -288 \zeta_3 &
\cofq{3}{292} & = 0 \\
\cofq{3}{293} & = 120 - 144 \zeta_3 &
\cofq{3}{294} & = -72 \zeta_3 &
\cofq{3}{295} & = 24 &
\cofq{3}{296} & = -48 \\
\cofq{3}{297} & = 48 &
\cofq{3}{298} & = 48 &
\cofq{3}{299} & = -96 &
\cofq{3}{300} & = -72 \\
\cofq{3}{301} & = -24 &
\cofq{3}{302} & = -48 &
\cofq{3}{303} & = 24 &
\cofq{3}{304} & = -\dfrac{21}{2} \\
\cofq{3}{305} & = 48 &
\cofq{3}{306} & = -24 &
\cofq{3}{307} & = -96 &
\cofq{3}{308} & = 48 \\
\cofq{3}{309} & = -12 &
\cofq{3}{310} & = -48 &
\cofq{3}{311} & = -18 &
\cofq{3}{312} & = -\dfrac{75}{2} \\
\cofq{3}{313} & = 12 &
\cofq{3}{314} & = 6 &
\cofq{3}{315} & = 12 
\end{align*}

\clearpage
\bibliographystyle{JHEP}
\bibliography{ref.bib}

\providecommand{\href}[2]{#2}\begingroup\raggedright\begin{thebibliography}{10}

\bibitem{Machacek:1983tz}
M.E.~Machacek and M.T.~Vaughn, \emph{{Two-loop renormalization group equations in a general quantum field theory: (I). Wave function renormalization}}, \href{https://doi.org/10.1016/0550-3213(83)90610-7}{\emph{Nucl. Phys.} {\bfseries B222} (1983) 83}.

\bibitem{Machacek:1983fi}
M.E.~Machacek and M.T.~Vaughn, \emph{{Two-loop renormalization group equations in a general quantum field theory (II). Yukawa couplings}}, \href{https://doi.org/10.1016/0550-3213(84)90533-9}{\emph{Nucl. Phys.} {\bfseries B236} (1984) 221}.

\bibitem{Machacek:1984zw}
M.E.~Machacek and M.T.~Vaughn, \emph{{Two-loop renormalization group equations in a general quantum field theory: (III). Scalar quartic couplings}}, \href{https://doi.org/10.1016/0550-3213(85)90040-9}{\emph{Nucl. Phys.} {\bfseries B249} (1985) 70}.

\bibitem{Jack:1984vj}
I.~Jack and H.~Osborn, \emph{{General Background Field Calculations With Fermion Fields}}, \href{https://doi.org/10.1016/0550-3213(85)90088-4}{\emph{Nucl. Phys. B} {\bfseries 249} (1985) 472}.

\bibitem{Pickering:2001aq}
A.~Pickering, J.~Gracey and D.~Jones, \emph{{Three loop gauge $\beta$-function for the most general single gauge-coupling theory}}, \href{https://doi.org/10.1016/S0370-2693(01)00624-4}{\emph{Phys. Lett. B} {\bfseries 510} (2001) 347} [\href{https://arxiv.org/abs/hep-ph/0104247}{{\ttfamily hep-ph/0104247}}].

\bibitem{Luo:2002ti}
M.~Luo, H.~Wang and Y.~Xiao, \emph{{Two-loop renormalization group equations in general gauge field theories}}, \href{https://doi.org/10.1103/PhysRevD.67.065019}{\emph{Phys. Rev. D} {\bfseries 67} (2003) 065019} [\href{https://arxiv.org/abs/hep-ph/0211440}{{\ttfamily hep-ph/0211440}}].

\bibitem{Mihaila:2012pz}
L.N.~Mihaila, J.~Salomon and M.~Steinhauser, \emph{{Renormalization constants and beta functions for the gauge couplings of the standard model to three-loop order}}, \href{https://doi.org/10.1103/PhysRevD.86.096008}{\emph{Phys. Rev. D} {\bfseries 86} (2012) 096008} [\href{https://arxiv.org/abs/1208.3357}{{\ttfamily 1208.3357}}].

\bibitem{Sperling:2013eva}
M.~Sperling, D.~Stöckinger and A.~Voigt, \emph{{Renormalization of vacuum expectation values in spontaneously broken gauge theories}}, \href{https://doi.org/10.1007/JHEP07(2013)132}{\emph{JHEP} {\bfseries 07} (2013) 132} [\href{https://arxiv.org/abs/1305.1548}{{\ttfamily 1305.1548}}].

\bibitem{Sperling:2013xqa}
M.~Sperling, D.~Stöckinger and A.~Voigt, \emph{{Renormalization of vacuum expectation values in spontaneously broken gauge theories: two-loop results}}, \href{https://doi.org/10.1007/JHEP01(2014)068}{\emph{JHEP} {\bfseries 01} (2014) 068} [\href{https://arxiv.org/abs/1310.7629}{{\ttfamily 1310.7629}}].

\bibitem{Mihaila:2014caa}
L.~Mihaila, \emph{{Three-loop gauge beta function in non-simple gauge groups}}, \href{https://doi.org/10.22323/1.197.0060}{\emph{PoS} {\bfseries RADCOR2013} (2013) 060}.

\bibitem{Schienbein:2018fsw}
I.~Schienbein, F.~Staub, T.~Steudtner and K.~Svirina, \emph{{Revisiting RGEs for general gauge theories}}, \href{https://doi.org/10.1016/j.nuclphysb.2018.12.001}{\emph{Nucl. Phys. B} {\bfseries 939} (2019) 1} [\href{https://arxiv.org/abs/1809.06797}{{\ttfamily 1809.06797}}].

\bibitem{Staub:2013tta}
F.~Staub, \emph{{SARAH 4 : A tool for (not only SUSY) model builders}}, \href{https://doi.org/10.1016/j.cpc.2014.02.018}{\emph{Comput. Phys. Commun.} {\bfseries 185} (2014) 1773} [\href{https://arxiv.org/abs/1309.7223}{{\ttfamily 1309.7223}}].

\bibitem{Litim:2020jvl}
D.F.~Litim and T.~Steudtner, \emph{{ARGES -- Advanced Renormalisation Group Equation Simplifier}},  \href{https://arxiv.org/abs/2012.12955}{{\ttfamily 2012.12955}}.

\bibitem{Sartore:2020gou}
L.~Sartore and I.~Schienbein, \emph{{PyR@TE 3}}, \href{https://doi.org/10.1016/j.cpc.2020.107819}{\emph{Comput. Phys. Commun.} {\bfseries 261} (2021) 107819} [\href{https://arxiv.org/abs/2007.12700}{{\ttfamily 2007.12700}}].

\bibitem{Thomsen:2021ncy}
A.E.~Thomsen, \emph{{Introducing RGBeta: a Mathematica package for the evaluation of renormalization group $ \beta $-functions}}, \href{https://doi.org/10.1140/epjc/s10052-021-09142-4}{\emph{Eur. Phys. J. C} {\bfseries 81} (2021) 408} [\href{https://arxiv.org/abs/2101.08265}{{\ttfamily 2101.08265}}].

\bibitem{Steudtner:FoRGEr}
T.~Steudtner, \emph{{FoRGEr}}, {\emph{{unpublished}} (2024) }.

\bibitem{tHooft:1973mfk}
G.~'t~Hooft, \emph{{Dimensional regularization and the renormalization group}}, \href{https://doi.org/10.1016/0550-3213(73)90376-3}{\emph{Nucl. Phys. B} {\bfseries 61} (1973) 455}.

\bibitem{Bardeen:1978yd}
W.A.~Bardeen, A.~Buras, D.~Duke and T.~Muta, \emph{{Deep Inelastic Scattering Beyond the Leading Order in Asymptotically Free Gauge Theories}}, \href{https://doi.org/10.1103/PhysRevD.18.3998}{\emph{Phys. Rev. D} {\bfseries 18} (1978) 3998}.

\bibitem{Poole:2019kcm}
C.~Poole and A.E.~Thomsen, \emph{{Constraints on 3- and 4-loop $\beta$-functions in a general four-dimensional Quantum Field Theory}}, \href{https://doi.org/10.1007/JHEP09(2019)055}{\emph{JHEP} {\bfseries 09} (2019) 055} [\href{https://arxiv.org/abs/1906.04625}{{\ttfamily 1906.04625}}].

\bibitem{Poole:2019txl}
C.~Poole and A.~Thomsen, \emph{{Weyl Consistency Conditions and $\gamma_5$}}, \href{https://doi.org/10.1103/PhysRevLett.123.041602}{\emph{Phys. Rev. Lett.} {\bfseries 123} (2019) 041602} [\href{https://arxiv.org/abs/1901.02749}{{\ttfamily 1901.02749}}].

\bibitem{Bednyakov:2021qxa}
A.~Bednyakov and A.~Pikelner, \emph{{Four-Loop Gauge and Three-Loop Yukawa Beta Functions in a General Renormalizable Theory}}, \href{https://doi.org/10.1103/PhysRevLett.127.041801}{\emph{Phys. Rev. Lett.} {\bfseries 127} (2021) 041801} [\href{https://arxiv.org/abs/2105.09918}{{\ttfamily 2105.09918}}].

\bibitem{Davies:2021mnc}
J.~Davies, F.~Herren and A.E.~Thomsen, \emph{{General gauge-Yukawa-quartic $\beta$-functions at 4-3-2-loop order}}, \href{https://doi.org/10.1007/JHEP01(2022)051}{\emph{JHEP} {\bfseries 01} (2022) 051} [\href{https://arxiv.org/abs/2110.05496}{{\ttfamily 2110.05496}}].

\bibitem{Jegerlehner:2000dz}
F.~Jegerlehner, \emph{{Facts of life with $\gamma_5$}}, \href{https://doi.org/10.1007/s100520100573}{\emph{Eur. Phys. J. C} {\bfseries 18} (2001) 673} [\href{https://arxiv.org/abs/hep-th/0005255}{{\ttfamily hep-th/0005255}}].

\bibitem{Bednyakov:2021ojn}
A.~Bednyakov and A.~Pikelner, \emph{{Six-loop beta functions in general scalar theory}}, \href{https://doi.org/10.1007/JHEP04(2021)233}{\emph{JHEP} {\bfseries 04} (2021) 233} [\href{https://arxiv.org/abs/2102.12832}{{\ttfamily 2102.12832}}].

\bibitem{Steudtner:2021fzs}
T.~Steudtner, \emph{{Towards general scalar-Yukawa renormalisation group equations at three-loop order}}, \href{https://doi.org/10.1007/JHEP05(2021)060}{\emph{JHEP} {\bfseries 05} (2021) 060} [\href{https://arxiv.org/abs/2101.05823}{{\ttfamily 2101.05823}}].

\bibitem{Jack:2023zjt}
I.~Jack, H.~Osborn and T.~Steudtner, \emph{{Explorations in Scalar Fermion Theories: $\beta$-functions, Supersymmetry and Fixed Points}},  \href{https://arxiv.org/abs/2301.10903}{{\ttfamily 2301.10903}}.

\bibitem{Chetyrkin:2012rz}
K.~Chetyrkin and M.~Zoller, \emph{{Three-loop $\beta$-functions for top-Yukawa and the Higgs self-interaction in the standard model}}, \href{https://doi.org/10.1007/JHEP06(2012)033}{\emph{JHEP} {\bfseries 06} (2012) 033} [\href{https://arxiv.org/abs/1205.2892}{{\ttfamily 1205.2892}}].

\bibitem{Litim:2023tym}
D.F.~Litim, N.~Riyaz, E.~Stamou and T.~Steudtner, \emph{{Asymptotic safety guaranteed at four-loop order}}, \href{https://doi.org/10.1103/PhysRevD.108.076006}{\emph{Phys. Rev. D} {\bfseries 108} (2023) 076006} [\href{https://arxiv.org/abs/2307.08747}{{\ttfamily 2307.08747}}].

\bibitem{Steudtner:2020tzo}
T.~Steudtner, \emph{{General scalar renormalisation group equations at three-loop order}}, \href{https://doi.org/10.1007/JHEP12(2020)012}{\emph{JHEP} {\bfseries 12} (2020) 012} [\href{https://arxiv.org/abs/2007.06591}{{\ttfamily 2007.06591}}].

\bibitem{Jack:2014pua}
I.~Jack and C.~Poole, \emph{{The a-function for gauge theories}}, \href{https://doi.org/10.1007/JHEP01(2015)138}{\emph{JHEP} {\bfseries 01} (2015) 138} [\href{https://arxiv.org/abs/1411.1301}{{\ttfamily 1411.1301}}].

\bibitem{Brod:2024zaz}
J.~Brod, L.~H\"udepohl, E.~Stamou and T.~Steudtner, \emph{{MaRTIn -- Manual for the ''Massive Recursive Tensor Integration''}},  \href{https://arxiv.org/abs/2401.04033}{{\ttfamily 2401.04033}}.

\bibitem{Misiak:1994zw}
M.~Misiak and M.~Munz, \emph{{Two loop mixing of dimension five flavor changing operators}}, \href{https://doi.org/10.1016/0370-2693(94)01553-O}{\emph{Phys. Lett. B} {\bfseries 344} (1995) 308} [\href{https://arxiv.org/abs/hep-ph/9409454}{{\ttfamily hep-ph/9409454}}].

\bibitem{Chetyrkin:1997fm}
K.G.~Chetyrkin, M.~Misiak and M.~Munz, \emph{{Beta functions and anomalous dimensions up to three loops}}, \href{https://doi.org/10.1016/S0550-3213(98)00122-9}{\emph{Nucl. Phys. B} {\bfseries 518} (1998) 473} [\href{https://arxiv.org/abs/hep-ph/9711266}{{\ttfamily hep-ph/9711266}}].

\bibitem{Bednyakov:2012en}
A.~Bednyakov, A.~Pikelner and V.~Velizhanin, \emph{{Yukawa coupling beta-functions in the Standard Model at three loops}}, \href{https://doi.org/10.1016/j.physletb.2013.04.038}{\emph{Phys. Lett. B} {\bfseries 722} (2013) 336} [\href{https://arxiv.org/abs/1212.6829}{{\ttfamily 1212.6829}}].

\bibitem{Bednyakov:2013eba}
A.~Bednyakov, A.~Pikelner and V.~Velizhanin, \emph{{Higgs self-coupling beta-function in the Standard Model at three loops}}, \href{https://doi.org/10.1016/j.nuclphysb.2013.07.015}{\emph{Nucl. Phys. B} {\bfseries 875} (2013) 552} [\href{https://arxiv.org/abs/1303.4364}{{\ttfamily 1303.4364}}].

\bibitem{Chetyrkin:2013wya}
K.~Chetyrkin and M.~Zoller, \emph{{$\beta$-function for the Higgs self-interaction in the Standard Model at three-loop level}}, \href{https://doi.org/10.1007/JHEP04(2013)091}{\emph{JHEP} {\bfseries 04} (2013) 091} [\href{https://arxiv.org/abs/1303.2890}{{\ttfamily 1303.2890}}].

\bibitem{Bednyakov:2013cpa}
A.~Bednyakov, A.~Pikelner and V.~Velizhanin, \emph{{Three-loop Higgs self-coupling beta-function in the Standard Model with complex Yukawa matrices}}, \href{https://doi.org/10.1016/j.nuclphysb.2013.12.012}{\emph{Nucl. Phys. B} {\bfseries 879} (2014) 256} [\href{https://arxiv.org/abs/1310.3806}{{\ttfamily 1310.3806}}].

\bibitem{Bednyakov:2014pia}
A.~Bednyakov, A.~Pikelner and V.~Velizhanin, \emph{{Three-loop SM beta-functions for matrix Yukawa couplings}}, \href{https://doi.org/10.1016/j.physletb.2014.08.049}{\emph{Phys. Lett. B} {\bfseries 737} (2014) 129} [\href{https://arxiv.org/abs/1406.7171}{{\ttfamily 1406.7171}}].

\bibitem{Ihrig:2019kfv}
B.~Ihrig, N.~Zerf, P.~Marquard, I.F.~Herbut and M.M.~Scherer, \emph{{Abelian Higgs model at four loops, fixed-point collision and deconfined criticality}}, \href{https://doi.org/10.1103/PhysRevB.100.134507}{\emph{Phys. Rev. B} {\bfseries 100} (2019) 134507} [\href{https://arxiv.org/abs/1907.08140}{{\ttfamily 1907.08140}}].

\bibitem{Zerf:2020mib}
N.~Zerf, R.~Boyack, P.~Marquard, J.A.~Gracey and J.~Maciejko, \emph{{Critical properties of the valence-bond-solid transition in lattice quantum electrodynamics}}, \href{https://doi.org/10.1103/PhysRevD.101.094505}{\emph{Phys. Rev. D} {\bfseries 101} (2020) 094505} [\href{https://arxiv.org/abs/2003.09226}{{\ttfamily 2003.09226}}].

\bibitem{Herren:2021yur}
F.~Herren and A.E.~Thomsen, \emph{{On ambiguities and divergences in perturbative renormalization group functions}}, \href{https://doi.org/10.1007/JHEP06(2021)116}{\emph{JHEP} {\bfseries 06} (2021) 116} [\href{https://arxiv.org/abs/2104.07037}{{\ttfamily 2104.07037}}].

\bibitem{Herren:2017uxn}
F.~Herren, L.~Mihaila and M.~Steinhauser, \emph{{Gauge and Yukawa coupling beta functions of two-Higgs-doublet models to three-loop order}}, \href{https://doi.org/10.1103/PhysRevD.97.015016}{\emph{Phys. Rev. D} {\bfseries 97} (2018) 015016} [\href{https://arxiv.org/abs/1712.06614}{{\ttfamily 1712.06614}}].

\bibitem{Fortin:2012cq}
J.-F.~Fortin, B.~Grinstein and A.~Stergiou, \emph{{Limit Cycles in Four Dimensions}}, \href{https://doi.org/10.1007/JHEP12(2012)112}{\emph{JHEP} {\bfseries 12} (2012) 112} [\href{https://arxiv.org/abs/1206.2921}{{\ttfamily 1206.2921}}].

\bibitem{Jack:1990eb}
I.~Jack and H.~Osborn, \emph{{Analogs of the $c$-theorem for four-dimensional renormalisable field theories}}, \href{https://doi.org/10.1016/0550-3213(90)90584-Z}{\emph{Nucl. Phys. B} {\bfseries 343} (1990) 647}.

\bibitem{Zerf:2017zqi}
N.~Zerf, L.N.~Mihaila, P.~Marquard, I.F.~Herbut and M.M.~Scherer, \emph{{Four-loop critical exponents for the Gross-Neveu-Yukawa models}}, \href{https://doi.org/10.1103/PhysRevD.96.096010}{\emph{Phys. Rev. D} {\bfseries 96} (2017) 096010} [\href{https://arxiv.org/abs/1709.05057}{{\ttfamily 1709.05057}}].

\bibitem{Mihaila:2017ble}
L.N.~Mihaila, N.~Zerf, B.~Ihrig, I.F.~Herbut and M.M.~Scherer, \emph{{Gross-Neveu-Yukawa model at three loops and Ising critical behavior of Dirac systems}}, \href{https://doi.org/10.1103/PhysRevB.96.165133}{\emph{Phys. Rev. B} {\bfseries 96} (2017) 165133} [\href{https://arxiv.org/abs/1703.08801}{{\ttfamily 1703.08801}}].

\bibitem{Jack:1996qq}
I.~Jack, D.~Jones and C.~North, \emph{{N=1 supersymmetry and the three loop anomalous dimension for the chiral superfield}}, \href{https://doi.org/10.1016/0550-3213(96)00269-6}{\emph{Nucl. Phys. B} {\bfseries 473} (1996) 308} [\href{https://arxiv.org/abs/hep-ph/9603386}{{\ttfamily hep-ph/9603386}}].

\bibitem{Parkes:1985hh}
A.~Parkes, \emph{{Three Loop Finiteness Conditions in $N=1$ Superyang-mills}}, \href{https://doi.org/10.1016/0370-2693(85)91357-7}{\emph{Phys. Lett. B} {\bfseries 156} (1985) 73}.

\bibitem{Fei:2016sgs}
L.~Fei, S.~Giombi, I.R.~Klebanov and G.~Tarnopolsky, \emph{{Yukawa CFTs and Emergent Supersymmetry}}, \href{https://doi.org/10.1093/ptep/ptw120}{\emph{PTEP} {\bfseries 2016} (2016) 12C105} [\href{https://arxiv.org/abs/1607.05316}{{\ttfamily 1607.05316}}].

\bibitem{Martin:1993zk}
S.P.~Martin and M.T.~Vaughn, \emph{{Two-loop renormalization group equations for soft supersymmetry-breaking couplings}}, \href{https://doi.org/10.1103/PhysRevD.50.2282}{\emph{Phys. Rev. D} {\bfseries 50} (1994) 2282} [\href{https://arxiv.org/abs/hep-ph/9311340}{{\ttfamily hep-ph/9311340}}].

\bibitem{Carmona:2021xtq}
A.~Carmona, A.~Lazopoulos, P.~Olgoso and J.~Santiago, \emph{{Matchmakereft: automated tree-level and one-loop matching}}, \href{https://doi.org/10.21468/SciPostPhys.12.6.198}{\emph{SciPost Phys.} {\bfseries 12} (2022) 198} [\href{https://arxiv.org/abs/2112.10787}{{\ttfamily 2112.10787}}].

\bibitem{Fuentes-Martin:2022jrf}
J.~Fuentes-Mart\'\i{}n, M.~K\"onig, J.~Pag\`es, A.E.~Thomsen and F.~Wilsch, \emph{{A proof of concept for matchete: an automated tool for matching effective theories}}, \href{https://doi.org/10.1140/epjc/s10052-023-11726-1}{\emph{Eur. Phys. J. C} {\bfseries 83} (2023) 662} [\href{https://arxiv.org/abs/2212.04510}{{\ttfamily 2212.04510}}].

\bibitem{Osborn:1989td}
H.~Osborn, \emph{{Derivation of a Four-dimensional $c$ Theorem}}, \href{https://doi.org/10.1016/0370-2693(89)90729-6}{\emph{Phys. Lett. B} {\bfseries 222} (1989) 97}.

\bibitem{Osborn:1991gm}
H.~Osborn, \emph{{Weyl consistency conditions and a local renormalisation group equation for general renormalisable field theories}}, \href{https://doi.org/10.1016/0550-3213(91)80030-P}{\emph{Nucl. Phys. B} {\bfseries 363} (1991) 486}.

\bibitem{tHooft:1972tcz}
G.~'t~Hooft and M.~Veltman, \emph{{Regularization and Renormalization of Gauge Fields}}, \href{https://doi.org/10.1016/0550-3213(72)90279-9}{\emph{Nucl. Phys. B} {\bfseries 44} (1972) 189}.

\bibitem{Breitenlohner:1977hr}
P.~Breitenlohner and D.~Maison, \emph{{Dimensional Renormalization and the Action Principle}}, \href{https://doi.org/10.1007/BF01609069}{\emph{Commun. Math. Phys.} {\bfseries 52} (1977) 11}.

\bibitem{Stockinger:2023ndm}
D.~St\"ockinger and M.~Wei\ss{}wange, \emph{{Full three-loop Renormalisation of an abelian chiral Gauge Theory with non-anticommuting $\gamma_5$ in the BMHV Scheme}},  \href{https://arxiv.org/abs/2312.11291}{{\ttfamily 2312.11291}}.

\end{thebibliography}\endgroup
\end{document}